\documentclass[11pt]{article}
\usepackage[top=1in, bottom=1in, left=1in, right=1in]{geometry}
\usepackage{tikz}
\usetikzlibrary{shapes,arrows,matrix,patterns,positioning,calc,snakes}
\usepackage{color}
\usepackage{verbatim}
\usepackage{graphicx}
\usepackage{algorithmic}
\usepackage{amssymb}
\usepackage{epstopdf}
\usepackage[linesnumbered,boxed,ruled,commentsnumbered]{algorithm2e}
\usepackage{float}
\usepackage{amsmath,amsthm,bm,color,epsfig,enumerate,caption}
\usepackage{hyperref}
\usepackage{booktabs}
\usepackage{multirow}
\usepackage{array}
\usepackage{tabu}
\usepackage{breqn}
\usepackage{mathtools}
\usepackage{multicol}
\usepackage{multirow}
\numberwithin{equation}{section}
 \usepackage{float}
\usepackage{lineno}

\newtheorem{theorem}{Theorem}[section]

\newtheorem{corollary}[theorem]{Corollary}

\newtheorem{lemma}[theorem]{Lemma}

\renewcommand{\appendix}[1]{
\section*{Appendix: #1}
}

\makeatletter
\newcommand*{\extendadd}{
  \mathbin{
    \mathpalette\extend@add{}
  }
}
\newcommand*{\extend@add}[2]{
  \ooalign{
    $\m@th#1\leftrightarrow$%
    \vphantom{$\m@th#1\updownarrow$}
    \cr
    \hfil$\m@th#1\updownarrow$\hfil
  }
}
\makeatother

\begin{document}

\title{Linear-Scaling Selected Inversion based on Hierarchical Interpolative Factorization for Self Green's Function for Modified Poisson-Boltzmann Equation in Two Dimensions}
\author{Yihui Tu$^1$, Qiyuan Pang$^2$, Haizhao Yang$^2$\footnote{Corresponding author.}, Zhenli Xu$^3$\footnote{Corresponding author.}  \\
$^1$ School of Mathematical Sciences, Shanghai Jiao Tong University, Shanghai 200240, China\\
$^2$ Department of Mathematics, Purdue University, West Lafayette, IN 47907, USA\\
$^3$ School of Mathematical Sciences, Institute of Natural Sciences and MoE-LSC, \\
Shanghai Jiao Tong University, Shanghai 200240, China
}

\date{\today}
\maketitle
\abstract{
This paper studies an efficient numerical method for solving modified Poisson-Boltzmann (MPB) equations with the self Green's function as a state equation
to describe electrostatic correlations in ionic systems. Previously, the most expensive point of the MPB solver is the evaluation of Green's function.
The evaluation of Green's function requires solving high-dimensional partial differential equations, which is the computational bottleneck for solving MPB equations.
Numerically, the MPB solver only requires the evaluation of Green's function as the diagonal part of the inverse of the discrete elliptic differential operator of the Debye-H\"uckel equation. Therefore, we develop a fast algorithm by a coupling of the selected inversion and hierarchical interpolative factorization. By the interpolative factorization, our new selected inverse algorithm achieves linear scaling to compute the diagonal of the inverse of this discrete operator. The accuracy and efficiency of the proposed algorithm will be demonstrated by extensive numerical results for solving MPB equations.}

\vspace{0.5cm}

\section{Introduction}
Electrostatic interaction plays important role in many fields of physical and biological sciences \cite{Schoch:RMP:08,daiguji2004ion,BKN+:PR:2005}, as well as materials science such as nanoparticle assembly \cite{Liljestrm2015ElectrostaticSO}. The Poisson-Boltzmann (PB) \cite{Gouy:JP:1910,Chapman:PM:1913} and Poisson-Nernst-Planck (PNP) equations \cite{PhysRevE.70.021506, Schuss2001DerivationOP} are often used to describe the electrostatic phenomena of equilibrium and dynamical systems, respectively. The PB equation is mean-field theory and fails to capture many-body properties such as dielectric variation and ion correlation, which are essential components of electrostatic behaviors of many systems. Many improved theories have been introduced in the literature to take into account these many-body effects \cite{modify1,Bazant2011DoubleLI, Eisenberg2020} and various numerical methods \cite{XML:PRE:2014,2014A,Liu2020APE} have been proposed to solve the systems efficiently. Among them, the Gaussian variational field theory \cite{pnp1,Podgornik1989ElectrostaticCF} is promising to describe the long-range Coulomb correlation including dielectric variation \cite{pnp2,Liu2018ModifiedPM,Ma2020ModifiedPM,Maggs2002}. The theory introduces the self-energy of a test ion as a correction to the mean-field potential energy, which is described by the self Green's function. Based on the self-energy,  the effect due to dielectric inhomogeneity has been considered \cite{modify3,Wang:PRE:2010,xu2,Ji2018AsymptoticAO}.
The Green's function used in the field theory satisfies the generalized Debye-H\"uckel (DH) equation for which the numerical solution is expensive due to its high spatial dimensions (including both source and field coordinates). By finite-difference discretization, the self Green's function corresponds to the diagonal of the inverse of the discrete elliptic differential operator of the DH equation.

In this paper, we propose a novel method for solving modified Poisson-Boltzmann (MPB) equations derived from the Gaussian variational field theory,  particularly, a fast algorithm for obtaining the  diagonal of the inverse matrix from the discretization of the DH equation. To show the basic idea, here we consider the following elliptic partial differential equation,

\begin{equation}
\label{eq:elpde1}
-\nabla \cdot(a(\bm{r}) \nabla u(\bm{r})) + b(\bm{r}) u(\bm{r}) = f(\bm{r}),\quad \bm{r} \in \boldsymbol{\Omega} \subset \mathbb{R}^{d}
\end{equation}
with an appropriate boundary condition, where $a(\bm{r})>0$,  $b(\bm{r})$, and $f(\bm{r})$ are functions on $\boldsymbol{\Omega}$, and $d = 2$. Then Eq. \eqref{eq:elpde1}  leads to  a linear system after finite-difference discretization,

\begin{equation*}\numberwithin{equation}{section}
\label{eq:dpde}
A u_{N} = f_{N},
\end{equation*}
where $A \in \mathbb{R}^{N\times N}$ is sparse, $u_{N}$ and $f_{N}$ are the discrete forms of $u(\bm{r})$ and $f(\bm{r})$, respectively. Our goal here is to compute the diagonal of $A^{-1}$ in $O(N)$ operations to obtain the self energy in the DH equation, which accelerates the numerical solver for MPB equations.

Determining the diagonal of a matrix inverse has been previously studied especially in electronic structure calculation based on sparsity and low-rankness, e.g., Lin {\it et al.}\cite{lin1,Lin2011:1,Lin2011:2} with $O(N^{3/2})$ computational complexity for 2D problems, and Xia {\it et al.} \cite{Xia2015} with $O(N \text{poly}(\log  N))$ complexity.
The selected inversion method \cite{lin1} applies a hierarchical decomposition of the computational domain $\boldsymbol{\Omega}$ and proposes a two-step procedure to form the diagonal of $A^{-1}$ with $O(N^{3/2})$ complexity for 2D problems. First, hierarchical Schur complements of the interior points for the blocks of the domain are constructed in a bottom-up pass. Second, the diagonal entries are extracted efficiently in a top-down pass by exploiting the hierarchical local dependence of the inverse matrices. The method in Refs. \cite{Lin2011:1,Lin2011:2} uses a supernode left-looking LDL factorization of $A$ to improve the efficiency of the selected inversion method by significantly reducing the prefactor in their complexity.  Structured multifrontal LDL factorizations \cite{Xia2015} are applied to obtain $O(N \text{poly}(\log  N))$ complexity.

Recently, hierarchical interpolative factorization (HIF) \cite{hifde} is used to a generalized LDL decomposition of $A$ within $O(N)$ complexity in 2D problems. The HIF is a fast approximation of Multifrontal Factorization (MF) by introducing additional levels of compression based on skeletonizing separator fronts.  Unlike \cite{Gillman2014ADS, Gillman2014AnOA, Grasedyck2007DomaindecompositionB, Schmitz2012AFD, Xia2009SuperfastMM} that keep the entire fronts but work with them implicitly using fast structured methods, the HIF allows us to reduce the fronts explicitly. Inspired by the HIF, we can replace the supernode left-looking LDL factorization with the HIF and revise the extraction procedure to approximate the diagonal of $A^{-1}$ within $O(N)$ operations for  2D problems. We will present the main idea of skeletonizing separator fronts in the HIF and its application to the selected inversion method with a visible example and a complexity estimation. For the detailed introduction to the selected inverse and the HIF, see Refs. \cite{Lin2011:1,Lin2011:2,hifde} and reference therein. We call the SelInvHIF for our algorithm in this paper.

The rest of the paper is organized as follows. Section \ref{NMmPBE} discusses iterative solvers for MPB equations. In Section \ref{SelInvHIF}, we introduce some preliminary tools of skeletonization of matrix factorization, then the details of SelInvHIF algorithm are presented.  Various numerical results of SelInvHIF are provided in Section \ref{Numerical} for solving the MPB equations. The conclusion and discussion for future work are presented in Section \ref{Conclusion}.

\section{Numerical Method for MPB Equations}
\label{NMmPBE}

In this section, we will present mathematical model and numerical scheme for the MPB equations to motivate our study of SelInvHIF. We consider an electrolyte of monovalent ions. In dimensionless units, the dynamics of the mobile ions can be described by the Nernst-Planck equations \cite{XML:PRE:2014},

\begin{equation} \label{NP}
\frac{\partial c_i}{\partial t}= \nabla\cdot D\left[\nabla c_i+c_i \nabla\left(z_i \Phi+ \Xi u\right)\right],
\end{equation}
where $c_i$ is the ionic concentration of the $i$th species, $z_i=\pm 1$ is the valence, and $D$ is the diffusion constant.
Here, $\Xi$ is the coupling parameter that describes the strength of the correlation energy. The Nernst-Planck equations are convection-diffusion equations, where the convection is due to
the electrostatic force on each ion, namely the gradient of the electrostatic energy.  In the modified PNP equations, the electrostatic energy is composed of the mean potential energy $z_i \Phi$ and the self energy $\Xi u$. The electric potential satisfies the Poisson equation,

\begin{equation*}
-\nabla \cdot \varepsilon \nabla \Phi=\rho_f+ \sum_i z_i c_i,
\end{equation*}
where $\varepsilon$ is the dielectric coefficient and $\rho_f$ is the fixed charge distribution. We suppose $\varepsilon=1$ in the electrolyte.
The self energy is represented by the self Green's function, described by the following DH equation,

\begin{equation*}
\begin{array}{c}
\left\{\begin{array}{l}
\displaystyle - \nabla \cdot \varepsilon \nabla G+ \sum_{i} z_{i}^{2} c_{i} G=  4\pi\delta\left(\bm{r}-\bm{r}^{\prime}\right), \\
\displaystyle u=\lim _{\bm{r}' \rightarrow \bm{r}}\left[G\left(\bm{r}, \bm{r}'\right)-G_{0}\left(\bm{r}, \bm{r}'\right)\right],
\end{array}\right.
\end{array}
\end{equation*}
where $G_{0}=1/(\varepsilon|\bm{r}-\bm{r}'|)$ is the free-space Green's function. It can be seen that the DH equation is coupled with the PNP equations as the ionic strength $I= \sum_{i} z_{i}^{2} c_{i}$ is determined by the Nernst-Planck equations. When the correlation effect can be ignored ($\Xi\rightarrow 0$), the whole systems become the classical PNP equations.

At equilibrium, the ionic flux in Eq. \eqref{NP} becomes zero and there is an explicit relation between the ionic concentration and the electrostatic energy,

\begin{equation*}
c_\pm=\frac{1}{2}\Lambda e^{\mp\Phi-\Xi u},
\end{equation*}
where $\Lambda$ is the fugacity determined by the far-field boundary conditions. The MPB equation can be obtained when the Boltzmann distributions are used in the Poisson equation, written as,

\begin{equation*}
-\nabla \cdot \varepsilon \nabla \Phi=\rho_f - \Lambda e^{\Xi u} \sinh \Phi.
\end{equation*}
Together with the DH equation, we have the following system of equations \cite{pnp1,pnp2},

\begin{equation}
\begin{array}{c}
\label{eq:PNP0}
\left\{\begin{array}{l}
\displaystyle -\nabla \cdot \varepsilon \nabla \Phi=\rho_f - \bm{\Lambda} e^{\Xi u} \sinh \Phi,\\
\displaystyle - \nabla \cdot \varepsilon \nabla G+ \bm{\Lambda} e^{\Xi u} G=4\pi \delta\left(\bm{r}-\bm{r}'\right), \\
\displaystyle u=\lim _{\bm{r}^{\prime} \rightarrow \bm{r}}\left[G\left(\bm{r}, \bm{r}^{\prime}\right)-G_{0}\left(\bm{r}, \bm{r}'\right)\right],
\end{array}\right.
\end{array}
\end{equation}
where the bold $\bm{\Lambda}$ indicates  that it is  $\Lambda$ in the electrolyte domain and zero outside.

Without loss of generality, we discuss the numerical method for solving Eq. \eqref{eq:PNP0} in this work. In particular, we focus on the numerical method for the self Green's function. The extension to the modified PNP is straightforward. A self-consistent iterative scheme for the solution of the partial differential equations in Eq. \eqref{eq:PNP0} was developed previously \cite{pb1}. The iterative scheme is the following,

\begin{equation}
\begin{array}{c}
\label{eq:PNP1}
\left\{\begin{array}{l}
\displaystyle -\nabla \cdot \varepsilon \nabla \Phi^{(k+1)} + \bm{\Lambda} e^{\Xi u^{(k)}} \sinh \Phi^{(k+1)} = \rho_f ,\\
\displaystyle - \nabla \cdot \varepsilon \nabla G^{(k+1)} + \bm{\Lambda} e^{\Xi u^{(k)}} G^{(k+1)} =4\pi \delta\left(\bm{r}-\bm{r}'\right), \\
\displaystyle u^{(k+1)} =\lim _{\bm{r}^' \rightarrow \bm{r}}\left[G^{(k+1)} \left(\bm{r}, \bm{r}^{\prime}\right)-G_{0}\left(\bm{r}, \bm{r}'\right)\right],
\end{array}\right.
\end{array}
\end{equation}
for $k=0,1, \cdots, M$. The stopping criteria is $\max \left|\Phi^{(M)}-\Phi^{(M-1)}\right|<\delta$ with a small error criteria $\delta$.

The iterative scheme consists of two alternating steps. One solves the first equation in Eq. \eqref{eq:PNP1} for $\Phi$ with given $u$. Then for the given $u$ and the acquired $\Phi$, one solves the second equation in Eq. \eqref{eq:PNP1} to obtain $G$ and then a new $u$ is computed via the third equation in Eq. \eqref{eq:PNP1}. These two steps are called PB and DH steps, respectively. One repeats these two steps until reaching the convergence criteria of the solution. Furthermore, the PB step can be efficiently solved using standard fast direct solvers. The problem at the core of obtaining Green's functions comes from the generalized DH equation.  In two dimensions, we can write the discretization of the DH equation by,

\begin{equation*}
\bm{A G}=\bm{E},
\end{equation*}
where $\bm{G}$  is a matrix representing the lattice Green's function, $\bm{E}$  is an identity matrix, and $\bm{A}$ is a coefficient matrix. Furthermore, we can arrive at the matrix inverse $\bm{G}=\bm{A}^{-1}$ directly to achieve the solution of the Green's function with expensive calculation.  To reduce the computation cost, let us express $U$ by

\begin{equation*}
U=\text{diag}\left(\bm{G}\right)-\text{diag} \left(\bm{G}_{\bm{0}}\right),
\end{equation*}
where $U$ is a vector representing the correlation function $u(r)$,  $\bm{G}_{\bm{0}}$ is a lattice Green's function in the free space, and $\operatorname{diag}(\cdot)$ is a vector representing the diagonals of the argument matrix. Thus,  calculating the whole inverse of the matrix directly is expensive and unnecessary. Our SelInvHIF is used to just obtain the diagonal entries of the operator matrix inverse to solve our target problem efficiently.

\section{The SelInvHIF Algorithm}
\label{SelInvHIF}
The SelInvHIF consists of two phases. In the first phase, we construct hierarchical Schur complements for the diagonal blocks of a matrix $A$ discretized uniformly from the differential operator in \eqref{eq:elpde1} on a rectangular domain $\boldsymbol{\Omega}$. In the second phase, the diagonal of the inverse of $A$ are extracted from the construction of the hierarchy of Schur complements. The total complexity of the proposed algorithm is analyzed at the end of this section. Before the formal introduction to our SelInvHIF algorithm, we first introduce some preliminary background of skeletonization of matrix factorization.

\subsection{Preliminaries}
\label{sec:skl}
Suppose $A$ is a symmetric matrix, $p$, $q$, $I$ and $J$ are index sets. $A_{pq}$ (or $A(I,J)$) denotes a submatrix of $A$ corresponding to rows in $p$ (or $I$) and columns in $q$ (or $J$). The notation ``$:$" is used to denote the whole row or column index set, e.g., $A_{:,q}$ consists of columns of $A$ corresponding to indices in $q$. In the discussion below, we will follow the same notation to denote submatrices.

 Suppose the differential operator in Eq. \eqref{eq:elpde1} is defined on a domain $\boldsymbol{\Omega}$. A typical discretization of a differential operator results in a sparse matrix with special structures. Let $A$ be a symmetric and nonsingular matrix
 
\begin{equation}\numberwithin{equation}{section}
\label{eq:A}
A = \left[\begin{array}{ccc}
A_{pp} & A_{qp}^{T} &  \\
A_{qp} & A_{qq} & A_{rq}^{T} \\
 & A_{rq} & A_{rr}
\end{array}\right]
\end{equation}
obtained from the discretization of the differential operator in Eq. \eqref{eq:elpde1}, where $p$, $q$, and $r$ are index sets of $A$ with a special order. In this matrix structure, we order rows and columns carefully such that $p$ is related to the degrees of freedom (DOFs) of the interior points of a small given domain $\mathcal{D}\subset \boldsymbol{\Omega}$,  $q$ corresponds to the DOFs on the boundary $\partial\mathcal{D}$, and $r$ is for the DOFs of the external domain $\boldsymbol{\Omega}/ \overline{\mathcal{D}}$. In general, the DOFs $q$ separates $p$ from $r$, which is often very large.

\subsubsection{Block Inversion}
The first preliminary tool we are going to use in SelInvHIF comes from the key observation in the selected inversion method: A diagonal block of the inverse of $A$ can be computed via a diagonal block of the inverse of a submatrix of $A$, the repeated application of which could lead to an efficient recursive algorithm to compute the diagonal of $A$. The key observation is based on Lemma \ref{lemma:1} below \cite{lin1}. Its proof is based on block Gaussian elimination.

\begin{lemma}
\label{lemma:1}
Suppose $A$ is given by (\ref{eq:A}) with a nonsingular $A_{pp}$ and
$G = A^{-1}$. Let $A_{1}$ be the Schur complement of $A_{pp}$, i.e.,

\begin{equation*}
\label{eq:A1}
A_1=\left[\begin{array}{cc}
A_{qq}-A_{qp}A_{pp}^{-1}A_{qp}^{T} &  A_{rq}^{T}\\
A_{rq} & A_{rr}\\
\end{array}\right],
\end{equation*}
and let $G_{1} = A_{1}^{-1}$.
Then it holds,

\begin{equation*}
\label{eq:Gpp}
G_{pp} = A_{pp}^{-1} + \begin{bmatrix} -A_{pp}^{-1}A_{qp}^T  ~~\boldsymbol{0}\end{bmatrix}
G_{1} \begin{bmatrix} -A_{pp}^{-1}A_{qp}^T ~~ \boldsymbol{0}\end{bmatrix} ^T,
\end{equation*}
where $G_{pp}$ is the submatrix of $G$ corresponding to the row and column index set $p$.
\end{lemma}
According to Lemma \ref{lemma:1}, the calculation of $G_{pp}$ only requires the values of $G_1$ associated with row and column indices in $q$, rather than the whole inverse of the Schur complement $A_1$. 
This implies that $G_{pp}$ is determined by $(G_{1})_{qq}=(A_1^{-1})_{qq}$. Similarly,  $(G_{1})_{qq}$ can be determined using a diagonal block of the inverse of the Schur complement of a submatrix of $A_1$. Repeatedly applying this idea results in a recursive algorithm to compute $G_{pp}$ efficiently.

\subsubsection{Interpolative Decomposition}
The second tool repeatedly applied in the SelInvHIF is the interpolative decomposition (ID) \cite{Cheng:SIAM:05} for low-rank matrices based on Lemma \ref{lemma:2} below.

\begin{lemma}
\label{lemma:2}
Let $A \in \mathbb{R}^{m\times n}$ with rank $k \leq \min(m,n)$ and $q$ be the set of all column indices of $A$. Then there exist a disjoint partition of $q = \hat{q}\cup \check{q}$ with $|\hat{q}|=k$ and a matrix $T_q \in \mathbb{R}^{k\times (n-k)}$ such that $A_{:,\check{q}} = A_{:,\hat{q}} T_{q}$.
\end{lemma}

The sets $\hat{q}$ and $\check{q}$  are called the skeleton and redundant indices, respectively. In particular, the redundant columns of $A$ can be expressed by its skeleton columns and the associated interpolation matrix from Lemma \ref{lemma:2}. The following corollary shows that  matrix $A$ can be sparsified by multiplying a triangular matrix constructed from the interpolation matrix $T_q$ in Lemma \ref{lemma:2}.

\begin{corollary}\label{cor:ID}
With the same assumptions and notations in Lemma \ref{lemma:2}, it holds that

\begin{equation*}
A\left[\begin{array}{cc}
I & \\
-T_{q} & I\\
\end{array}\right]=
\left[\begin{array}{cc}A_{:,\check{q}} & A_{:,\hat{q}}\\
\end{array}\right]
\left[\begin{array}{cc}
I & \\
-T_{q} & I\\
\end{array}\right]
=
\left[\begin{array}{cc}
\boldsymbol{0} & A_{:,\hat{q}}\\
\end{array}\right].
\end{equation*}
\end{corollary}

\subsubsection{Block Inversion with Skeletonization}
The application of Corollary  \ref{cor:ID} can eliminate redundant DOFs of a dense matrix with low-rank off-diagonal blocks to form a structured matrix of the form \eqref{eq:A} such that we can apply Lemma \ref{lemma:1}. This idea is called block inversion with skeletonization summarized in Lemma \ref{lemma:4} below. The skeletonization idea was originally proposed in the HIF \cite{hifde}.

\begin{lemma}
\label{lemma:4}
Let

\begin{equation*}\numberwithin{equation}{section}
A =
\left[\begin{array}{cc}
A_{pp} & A_{qp}^{T}\\
A_{qp} & A_{qq}
\end{array}\right]
\end{equation*}
be symmetric with $A_{qp}$ low-rank. Let $p = \hat{p}\cup \check{p}$ and $T_{p}$ satisfy $A_{q\check{p}} = A_{q\hat{p}} T_{p}$. Without loss of generality, rewrite

\begin{equation*}\numberwithin{equation}{section}
A = \left[\begin{array}{ccc}
A_{\check{p}\check{p}} & A_{\hat{p}\check{p}}^{T} & A_{q\check{p}}^{T} \\
A_{\hat{p}\check{p}} & A_{\hat{p}\hat{p}} & A_{q\hat{p}}^{T} \\
A_{q\check{p}} & A_{q\hat{p}} & A_{qq}
\end{array}\right]
\end{equation*}
and define

\begin{equation*}\numberwithin{equation}{section}
Q_{p} =
\left[\begin{array}{ccc}
I & & \\
-T_{p} & I & \\
& & I
\end{array}\right].
\end{equation*}
Then

\begin{equation}\label{eq:Ab}
\numberwithin{equation}{section}
\bar{A} \triangleq Q_{p}^{T}A Q_{p} =
\left[\begin{array}{ccc}
B_{\check{p}\check{p}} & B_{\hat{p}\check{p}}^{T} &  \\
B_{\hat{p}\check{p}} & A_{\hat{p}\hat{p}} & A_{q\hat{p}}^{T} \\
 & A_{q\hat{p}} & A_{qq}
\end{array}\right],
\end{equation}
where

\begin{equation*}\numberwithin{equation}{section}
B_{\check{p}\check{p}} = A_{\check{p}\check{p}}-T_{p}^{T}A_{\hat{p}\check{p}}-A_{\hat{p}\check{p}}^{T}T_{p}+T_{p}^{T}A_{\hat{p}\hat{p}}T_{p},
~\hbox{and}~
B_{\hat{p}\check{p}} = A_{\hat{p}\check{p}} - A_{\hat{p}\hat{p}}T_{p}.
\end{equation*}
Assume that $B_{\check{p}\check{p}}$ is nonsingular. Let $G = A^{-1}$, $\bar{G} = \bar{A}^{-1}$, $G_{1} = G_{\hat{p}\cup q, \hat{p}\cup q}$, $\bar{A}_{1}$ be the Schur complement of  $B_{\check{p}\check{p}}$, i.e.,

\begin{equation*}\numberwithin{equation}{section}
\label{eq:Schur1}
\bar{A}_{1}=
\left[\begin{array}{cc}
A_{\hat{p}\hat{p}}-B_{\hat{p}\check{p}}B_{\check{p}\check{p}}^{-1}B_{\hat{p}\check{p}}^{T} &  A_{q\hat{p}}^{T}\\
A_{q\hat{p}} & A_{qq}\\
\end{array}\right],
\end{equation*}
and $\bar{G}_{1} = \bar{A}_{1}^{-1}$. Then the following formulas hold by Lemma \ref{lemma:1} and \eqref{eq:Ab},

\begin{equation*}\numberwithin{equation}{section}
G_{\check{p}\check{p}} = \bar{G}_{\check{p}\check{p}} = B_{\check{p}\check{p}}^{-1}+\begin{bmatrix} -B_{\check{p}\check{p}}^{-1}B_{\hat{p}\check{p}}^{T}\ \boldsymbol{0} \end{bmatrix}\bar{G}_{1}\begin{bmatrix} -B_{\check{p}\check{p}}^{-1}B_{\hat{p}\check{p}}^{T}\ \boldsymbol{0} \end{bmatrix}^T,
\end{equation*}

\begin{equation*}\numberwithin{equation}{section}
G_{1} =
\left[\begin{array}{cc}
T_{p}B_{\check{p}\check{p}}^{-1}T_{p}^{T} & \\
& \boldsymbol{0}
\end{array}\right]+
\left[\begin{array}{cc}
T_{p}B_{\check{p}\check{p}}^{-1}B_{\hat{p}\check{p}}^{T}+I & \\
& I
\end{array}\right]\bar{G}_{1}
\left[\begin{array}{cc}
B_{\hat{p}\check{p}}B_{\check{p}\check{p}}^{-1}T_{p}^{T}+I & \\
& I
\end{array}\right].
\end{equation*}
 \end{lemma}

 According to Lemma \ref{lemma:4}, the calculation of $G_{\check{p}\check{p}}$ only requires the values of $\bar{G}_{1}$ associated with row and column indices in $\hat{p}$, rather than the whole inverse of the Schur complement, i.e., this observation implies that $G_{\check{p}\check{p}}$ is determined by  $(\bar{G}_1)_{\hat{p}\hat{p}}$, a diagonal block of the inverse of the matrix $\bar{A}_1$ of a smaller size than the original larger matrix $A$.  Though $\bar{A}_1$ might be dense, as long as it has low-rank off-diagonal blocks, the same idea as in  \eqref{eq:Ab} can be applied to $\bar{A}_1$ to compute a diagonal block of the inverse of $\bar{A}_1$, which forms a recursive algorithm to compute the diagonal blocks of $A$ efficiently.

This skeletonization is the key contribution of Ho and Ying \cite{hifde} and the reason why it is applicable is stated as follows. The key conclusion is that the above Schur complements have specific low-rank structures. The matrix $A_{pp}^{-1}$ from a local differential operator often has numerically low-rank off-diagonal blocks. Especially,  the  Schur complement interaction $A_{qq}-A_{qp}A_{pp}^{-1}A_{qp}^{T}$  also has the same rank structure, which is verified by numerical experiments. In the next subsection, we apply Lemma \ref{lemma:1} and  Lemma \ref{lemma:4} to construct the hierarchical Schur complements for the diagonal blocks of a matrix $A$.

\subsection{Hierarchy of Schur Complements}

We discuss the differential operator domain with a grid size $\sqrt{N}\times \sqrt{N} = r_0 2^{L-1}\times r_0 2^{L-1}$ and an initial index set $J_{0}$ (e.g., row-major ordering). The corresponding matrix $A$ is of size $N\times N$.  A hierarchical disjoint partition of the domain  $\boldsymbol{\Omega}$ (bipartition in each dimension) is performed with $r_0\times r_0$ as the size of leaf domains and $L$ as the total number of integer levels. Between $L$ integer levels, $L-1$ fractional levels are designed to take advantage of the low-rankness of $A$ as much as possible. The hierarchy construction of Schur complements will be conducted at levels $1$, $\frac{3}{2}$, $2$, $\frac{5}{2}$, $\dots$, and $L$.

\begin{figure}[htp]
\begin{minipage}{\textwidth}
\centering
\resizebox{12.0cm}{!}{
\input{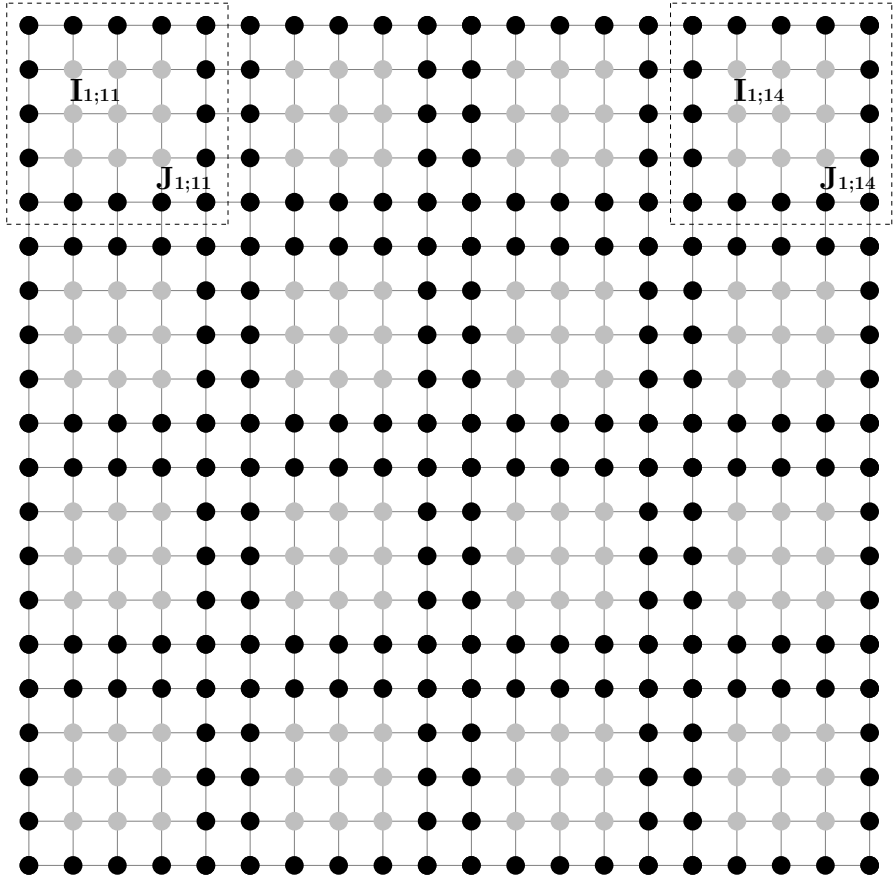}
}

\end{minipage}
\caption{The DOFs in the first level.  The domain is partitioned into 16 blocks in the Level 1 and the dash lines show two of blocks in the decomposition. In this and following integral levels,  the interior points are marked in gray and the boundary points are marked in black. It's noted that the blocks share edge in practice to reduce the prefactor.}
\label{fig:lel1-0}
\end{figure}

For simplicity, consider the case of $r_0=5$ and $L=3$. The whole domain is considered as the top level (Level $3$) and is divided into four blocks at the next level (Level $2$). Each block is again partitioned into four sub-blocks at a lower level (Level $1$). Between two adjacent integer levels, one fractional level will be added to skeletonize low-rank matrices corresponding to the fronts between domain blocks. Hence, the whole domain is divided into $2^{L-1}\times 2^{L-1}=16$ blocks at the bottom level (Level 1) as shown in Figure \ref{fig:lel1-0}.

\subsubsection{Level $\ell=1$} 
The domain  $\boldsymbol{\Omega}$ is partitioned into $2^{L-\ell}\times 2^{L-\ell}=4\times 4$ disjoint blocks at Level $\ell=1$. All points are separated into interior points and boundary points in each block. The interior points indicate that they are not related to the points of other blocks, and the boundary points indicate that they are related to the points of other blocks, i.e., having neighboring points in other blocks. The index set of the interior points are denoted as $I_{1;ij}$ for each block (gray points in Figure \ref{fig:lel1-0}), and
the index set of the boundary points are denoted as $J_{1;ij}$  of each block (black points in Figure \ref{fig:lel1-0}), where $i,j=1,2,3,4$ are the indices of blocks in each dimension. The locality of differential operators leads to  $A(I_{1;ij},I_{1;i'j'}) = 0$ (or $A(I_{1;ij},J_{1;i'j'}) = 0$) if $(i,j) \neq (i',j')$.

Firstly, Gauss elimination is used to eliminate the interior points, thereby shifting the focus of the problem to the boundary points. To do this, we need to apply necessary row and column permutations to the matrix $A$ defined with the index set $J_ {0}$ such that all the interior points are in front of the boundary points, that is, the indices $J_{0}$ is changed to

\begin{equation*}\numberwithin{equation}{section}
J_{0} \stackrel{P_{1}}{\longrightarrow} (I_{1;11}I_{1;12}...I_{1;44}|J_{1;11}J_{1;12}...J_{1;44})
\end{equation*}
by a permutation matrix $P_1$. Then all interior points and boundary points are separated by notation $|$. In fact, the permutation matrix $P_{1}$ can permute the matrix $A$ into a new matrix

\begin{equation*}\numberwithin{equation}{section}
A_{1} = P_{1}^{-1}A P_{1}
\end{equation*}
with index set $(I_{1}|J_{1})$, where $I_{1} = I_{1;11}I_{1;12}...I_{1;44}$ gathering all the interior points, and $J_{1}=J_{1;11}J_{1;12}...J_{1;44}$ for all the boundary points. Represent $A_{1}$ via

\begin{equation}\numberwithin{equation}{section}\label{A1}
A_{1} = P_{1}^{-1}A P_{1} =
\left[\begin{array}{cc}
U_{1} & V_{1}^{T}\\
V_{1} & W_{1}
\end{array}\right],
\end{equation}
where $U_{1}$ is a block diagonal matrix as follows because the interior points of different blocks in Figure \ref{fig:lel1-0} are not related to the points of other blocks:

\begin{equation*}\numberwithin{equation}{section}
U_{1} = A_{1}(I_{1},I_{1})=
\left[\begin{array}{cccc}
U_{1;11}& & & \\
& U_{1;12} & & \\
& & \ddots & \\
& & & U_{1;44}
\end{array}\right]
\end{equation*}
with $U_{1;ij} = A_{1}(I_{1;ij},I_{1;ij}).$ Moreover, $V_ {1}$ is also a block diagonal matrix as follows because the interior points of each block just is related to the boundary points of the same block:

\begin{equation*}\numberwithin{equation}{section}
V_{1} = A_{1}(J_{1},I_{1}) =
\left[\begin{array}{cccc}
V_{1;11} & & & \\
& V_{1;12} & & \\
& & \ddots &  \\
& & & V_{1;44}
\end{array}\right]
\end{equation*}
with $V_{1;ij} = A_{1}(J_{1;ij},I_{1;ij}).$
As for $W_{1}$, we just write it with index set,

\begin{equation*}\numberwithin{equation}{section}
W_{1} = A_{1}(J_{1},J_{1}).
\end{equation*}
The inverse of $U_{1}$ can be computed directly as follows since it is a block diagonal matrix with diagonal blocks of a small size $(r_0-2)^2 \times (r_0-2)^2$,

\begin{equation*}\numberwithin{equation}{section}
U_{1}^{-1} =
\left[\begin{array}{cccc}
U_{1;11}^{-1}& & & \\
& U_{1;12}^{-1} & & \\
& & \ddots & \\
& & & U_{1;44}^{-1}
\end{array}\right].
\end{equation*}
Therefore,  we can obtain the following inverse by Gaussian elimination,

\begin{equation}\numberwithin{equation}{section}
\label{eq:A1-1}
A_{1}^{-1} =
\left[\begin{array}{cc}
U_{1} & V_{1}^{T}\\
V_{1} & W_{1}
\end{array}\right]^{-1}=
L_{1}^{T}
\left[\begin{array}{cc}
U_{1}^{-1} & \\
 & (W_{1}-V_{1}U_{1}^{-1}V_{1}^{T})^{-1}
\end{array}\right]L_{1}
\end{equation}
with $
L_{1} =
\left[\begin{array}{cc}
I & \\
-V_{1}U_{1}^{-1} & I
\end{array}\right].
$ Furthermore, $V_{1}U_{1}^{-1} $ can  be computed independently within each block with block diagonal matrices $V_{1}$ and $U_{1}^{-1}$,

\begin{equation*}\numberwithin{equation}{section}
V_{1}U_{1}^{-1} =
\left[\begin{array}{cccc}
V_{1;11}U_{1;11}^{-1} & & & \\
& V_{1;12}U_{1;12}^{-1} & & \\
& & \ddots & \\
& & & V_{1;44}U_{1;44}^{-1}
\end{array}\right].
\end{equation*}
Moreover, the block diagonal matrix $V_{1}U_{1}^{-1}V_{1}^{T}$  also can be represented as

\begin{equation*}\numberwithin{equation}{section}
V_{1}U_{1}^{-1}V_{1}^{T} =
\left[\begin{array}{cccc}
V_{1;11}U_{1;11}^{-1}V_{1;11}^{T} & & & \\
& V_{1;12}U_{1;12}^{-1}V_{1;12}^{T} & & \\
& & \ddots & \\
& & & V_{1;44}U_{1;44}^{-1}V_{1;44}^{T}
\end{array}\right].
\end{equation*}

Combining (\ref{A1}) and (\ref{eq:A1-1}), we have

\begin{equation}\numberwithin{equation}{section}
G = A^{-1} =  P_{1}A_{1}^{-1}P_{1}^{-1} = P_{1}L_{1}^{T}
\left[\begin{array}{cc}
U_{1}^{-1} & \\
& G_{1}
\end{array}\right]L_{1}P_{1}^{-1},
\label{eqn:G1}
\end{equation}
where $G_{1} = (W_{1}-V_{1}U_{1}^{-1}V_{1}^{T})^{-1}$ is the inverse of the Schur complement of $U_{1}$. Therefore, we reduce the problem to a smaller matrix  $W_{1}-V_{1}U_{1}^{-1}V_{1}^{T}$ by eliminating interior points, which is essentially the block inversion idea in Lemma \ref{lemma:1}.

\begin{figure}[htp]
\begin{minipage}{\textwidth}
\centering
\resizebox{12.0cm}{!}{
\input{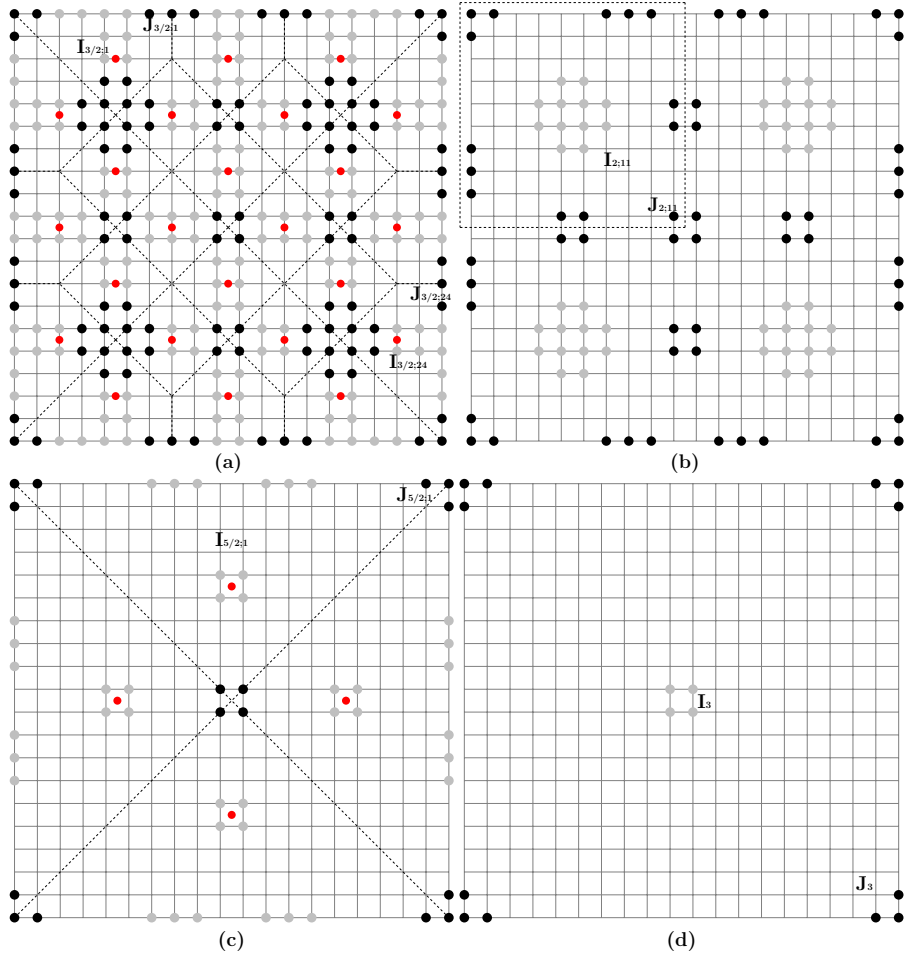}
\begin{tikzpicture}

\draw [help lines] (0,0) grid (19,19);

\fill [black] ($(0,0)$) circle (6pt);
\fill [black] ($(1,0)$) circle (6pt);
\fill [black] ($(6,0)$) circle (6pt);
\fill [black] ($(7,0)$) circle (6pt);
\fill [black] ($(8,0)$) circle (6pt);
\fill [black] ($(11,0)$) circle (6pt);
\fill [black] ($(12,0)$) circle (6pt);
\fill [black] ($(13,0)$) circle (6pt);
\fill [black] ($(18,0)$) circle (6pt);
\fill [black] ($(19,0)$) circle (6pt);

\fill [black] ($(0,1)$) circle (6pt);
\fill [black] ($(19,1)$) circle (6pt);

\fill [lightgray] ($(4,3)$) circle (6pt);
\fill [lightgray] ($(5,3)$) circle (6pt);
\fill [lightgray] ($(14,3)$) circle (6pt);
\fill [lightgray] ($(15,3)$) circle (6pt);

\fill [lightgray] ($(3,4)$) circle (6pt);
\fill [lightgray] ($(4,4)$) circle (6pt);
\fill [lightgray] ($(5,4)$) circle (6pt);
\fill [lightgray] ($(6,4)$) circle (6pt);
\fill [black] ($(9,4)$) circle (6pt);
\fill [black] ($(10,4)$) circle (6pt);
\fill [lightgray] ($(13,4)$) circle (6pt);
\fill [lightgray] ($(14,4)$) circle (6pt);
\fill [lightgray] ($(15,4)$) circle (6pt);
\fill [lightgray] ($(16,4)$) circle (6pt);

\fill [lightgray] ($(3,5)$) circle (6pt);
\fill [lightgray] ($(4,5)$) circle (6pt);
\fill [lightgray] ($(5,5)$) circle (6pt);
\fill [lightgray] ($(6,5)$) circle (6pt);
\fill [black] ($(9,5)$) circle (6pt);
\fill [black] ($(10,5)$) circle (6pt);
\fill [lightgray] ($(13,5)$) circle (6pt);
\fill [lightgray] ($(14,5)$) circle (6pt);
\fill [lightgray] ($(15,5)$) circle (6pt);
\fill [lightgray] ($(16,5)$) circle (6pt);

\fill [black] ($(0,6)$) circle (6pt);
\fill [lightgray] ($(4,6)$) circle (6pt);
\fill [lightgray] ($(5,6)$) circle (6pt);
\fill [lightgray] ($(14,6)$) circle (6pt);
\fill [lightgray] ($(15,6)$) circle (6pt);
\fill [black] ($(19,6)$) circle (6pt);

\fill [black] ($(0,7)$) circle (6pt);
\fill [black] ($(19,7)$) circle (6pt);

\fill [black] ($(0,8)$) circle (6pt);
\fill [black] ($(19,8)$) circle (6pt);

\fill [black] ($(4,9)$) circle (6pt);
\fill [black] ($(5,9)$) circle (6pt);
\fill [black] ($(9,9)$) circle (6pt);
\fill [black] ($(10,9)$) circle (6pt);
\fill [black] ($(14,9)$) circle (6pt);
\fill [black] ($(15,9)$) circle (6pt);

\fill [black] ($(0,19)$) circle (6pt);
\fill [black] ($(1,19)$) circle (6pt);
\fill [black] ($(6,19)$) circle (6pt);
\fill [black] ($(7,19)$) circle (6pt);
\fill [black] ($(8,19)$) circle (6pt);
\fill [black] ($(11,19)$) circle (6pt);
\fill [black] ($(12,19)$) circle (6pt);
\fill [black] ($(13,19)$) circle (6pt);
\fill [black] ($(18,19)$) circle (6pt);
\fill [black] ($(19,19)$) circle (6pt);

\fill [black] ($(0,18)$) circle (6pt);
\fill [black] ($(19,18)$) circle (6pt);

\fill [lightgray] ($(4,16)$) circle (6pt);
\fill [lightgray] ($(5,16)$) circle (6pt);
\fill [lightgray] ($(14,16)$) circle (6pt);
\fill [lightgray] ($(15,16)$) circle (6pt);

\fill [lightgray] ($(3,15)$) circle (6pt);
\fill [lightgray] ($(4,15)$) circle (6pt);
\fill [lightgray] ($(5,15)$) circle (6pt);
\fill [lightgray] ($(6,15)$) circle (6pt);
\fill [black] ($(9,15)$) circle (6pt);
\fill [black] ($(10,15)$) circle (6pt);
\fill [lightgray] ($(13,15)$) circle (6pt);
\fill [lightgray] ($(14,15)$) circle (6pt);
\fill [lightgray] ($(15,15)$) circle (6pt);
\fill [lightgray] ($(16,15)$) circle (6pt);

\fill [lightgray] ($(3,14)$) circle (6pt);
\fill [lightgray] ($(4,14)$) circle (6pt);
\fill [lightgray] ($(5,14)$) circle (6pt);
\fill [lightgray] ($(6,14)$) circle (6pt);
\fill [black] ($(9,14)$) circle (6pt);
\fill [black] ($(10,14)$) circle (6pt);
\fill [lightgray] ($(13,14)$) circle (6pt);
\fill [lightgray] ($(14,14)$) circle (6pt);
\fill [lightgray] ($(15,14)$) circle (6pt);
\fill [lightgray] ($(16,14)$) circle (6pt);

\fill [black] ($(0,13)$) circle (6pt);
\fill [lightgray] ($(4,13)$) circle (6pt);
\fill [lightgray] ($(5,13)$) circle (6pt);
\fill [lightgray] ($(14,13)$) circle (6pt);
\fill [lightgray] ($(15,13)$) circle (6pt);
\fill [black] ($(19,13)$) circle (6pt);

\fill [black] ($(0,12)$) circle (6pt);
\fill [black] ($(19,12)$) circle (6pt);

\fill [black] ($(0,11)$) circle (6pt);
\fill [black] ($(19,11)$) circle (6pt);

\fill [black] ($(4,10)$) circle (6pt);
\fill [black] ($(5,10)$) circle (6pt);
\fill [black] ($(9,10)$) circle (6pt);
\fill [black] ($(10,10)$) circle (6pt);
\fill [black] ($(14,10)$) circle (6pt);
\fill [black] ($(15,10)$) circle (6pt);

\draw [dashed] (-0.5,9.5) to (9.5,9.5);
\draw [dashed] (9.5,9.5) to (9.5,19.5);
\draw [dashed] (9.5,19.5) to (-0.5,19.5);
\draw [dashed] (-0.5,19.5) to (-0.5,9.5);
\node (a) at (6.5,12.5) {\textbf{\Huge I\large{2;11}}};
\node (a) at (8.5,10.5) {\textbf{\Huge J\large{2;11}}};

\node (a) at (9.5,-1) {\textbf{\Huge (b) }};

\end{tikzpicture}

}\\

\resizebox{12.0cm}{!}{

\begin{tikzpicture}

\draw [help lines] (0,0) grid (19,19);

\fill [black] ($(0,0)$) circle (6pt);
\fill [black] ($(1,0)$) circle (6pt);
\fill [lightgray] ($(6,0)$) circle (6pt);
\fill [lightgray] ($(7,0)$) circle (6pt);
\fill [lightgray] ($(8,0)$) circle (6pt);
\fill [lightgray] ($(11,0)$) circle (6pt);
\fill [lightgray] ($(12,0)$) circle (6pt);
\fill [lightgray] ($(13,0)$) circle (6pt);
\fill [black] ($(18,0)$) circle (6pt);
\fill [black] ($(19,0)$) circle (6pt);

\fill [black] ($(0,1)$) circle (6pt);
\fill [black] ($(19,1)$) circle (6pt);

\fill [lightgray] ($(9,4)$) circle (6pt);
\fill [lightgray] ($(10,4)$) circle (6pt);

\fill [lightgray] ($(9,5)$) circle (6pt);
\fill [lightgray] ($(10,5)$) circle (6pt);

\fill [lightgray] ($(0,6)$) circle (6pt);
\fill [lightgray] ($(19,6)$) circle (6pt);

\fill [lightgray] ($(0,7)$) circle (6pt);
\fill [lightgray] ($(19,7)$) circle (6pt);

\fill [lightgray] ($(0,8)$) circle (6pt);
\fill [lightgray] ($(19,8)$) circle (6pt);

\fill [lightgray] ($(4,9)$) circle (6pt);
\fill [lightgray] ($(5,9)$) circle (6pt);
\fill [black] ($(9,9)$) circle (6pt);
\fill [black] ($(10,9)$) circle (6pt);
\fill [lightgray] ($(14,9)$) circle (6pt);
\fill [lightgray] ($(15,9)$) circle (6pt);

\fill [black] ($(0,19)$) circle (6pt);
\fill [black] ($(1,19)$) circle (6pt);
\fill [lightgray] ($(6,19)$) circle (6pt);
\fill [lightgray] ($(7,19)$) circle (6pt);
\fill [lightgray] ($(8,19)$) circle (6pt);
\fill [lightgray] ($(11,19)$) circle (6pt);
\fill [lightgray] ($(12,19)$) circle (6pt);
\fill [lightgray] ($(13,19)$) circle (6pt);
\fill [black] ($(18,19)$) circle (6pt);
\fill [black] ($(19,19)$) circle (6pt);

\fill [black] ($(0,18)$) circle (6pt);
\fill [black] ($(19,18)$) circle (6pt);

\fill [lightgray] ($(9,15)$) circle (6pt);
\fill [lightgray] ($(10,15)$) circle (6pt);

\fill [lightgray] ($(9,14)$) circle (6pt);
\fill [lightgray] ($(10,14)$) circle (6pt);

\fill [lightgray] ($(0,13)$) circle (6pt);
\fill [lightgray] ($(19,13)$) circle (6pt);

\fill [lightgray] ($(0,12)$) circle (6pt);
\fill [lightgray] ($(19,12)$) circle (6pt);

\fill [lightgray] ($(0,11)$) circle (6pt);
\fill [lightgray] ($(19,11)$) circle (6pt);

\fill [lightgray] ($(4,10)$) circle (6pt);
\fill [lightgray] ($(5,10)$) circle (6pt);
\fill [black] ($(9,10)$) circle (6pt);
\fill [black] ($(10,10)$) circle (6pt);
\fill [lightgray] ($(14,10)$) circle (6pt);
\fill [lightgray] ($(15,10)$) circle (6pt);

\fill [red] ($(9.5,4.5)$) circle (5pt);
\fill [red] ($(9.5,14.5)$) circle (5pt);
\fill [red] ($(4.5,9.5)$) circle (5pt);
\fill [red] ($(14.5,9.5)$) circle (5pt);

\draw [dashed] (0,0) to (19,19);
\draw [dashed] (0,19) to (19,0);

\node (a) at (9.5,16.5){\textbf{\Huge I\large{5/2;1}}} ;
\node (a) at (17.5,18.5){\textbf{\Huge J\large{5/2;1}}};

\node (a) at (9.5,-1){\textbf{\Huge (c)}};
\end{tikzpicture}
\begin{tikzpicture}

\draw [help lines] (0,0) grid (19,19);

\fill [black] ($(0,0)$) circle (6pt);
\fill [black] ($(1,0)$) circle (6pt);
\fill [black] ($(18,0)$) circle (6pt);
\fill [black] ($(19,0)$) circle (6pt);

\fill [black] ($(0,1)$) circle (6pt);
\fill [black] ($(19,1)$) circle (6pt);

\fill [lightgray] ($(9,9)$) circle (6pt);
\fill [lightgray] ($(10,9)$) circle (6pt);

\fill [black] ($(0,19)$) circle (6pt);
\fill [black] ($(1,19)$) circle (6pt);
\fill [black] ($(18,19)$) circle (6pt);
\fill [black] ($(19,19)$) circle (6pt);

\fill [black] ($(0,18)$) circle (6pt);
\fill [black] ($(19,18)$) circle (6pt);

\fill [lightgray] ($(9,10)$) circle (6pt);
\fill [lightgray] ($(10,10)$) circle (6pt);

\node (a) at (10.5,9.5) {\textbf{\Huge I\large{3}}};
\node (a) at (17.5,1.5){\textbf{\Huge J\large{3}}};

\node (a) at (9.5,-1){\textbf{\Huge (d)}};
\end{tikzpicture}
}

\end{minipage}
\caption{
(a): The DOFs in the Level 3/2. The domain is partitioned into 24 Voronoi cells about the edge centers. In this and following fractional level, the redundant DOFs are marked in gray and the skeleton DOFs are marked in black.
(b): The DOFs in the Level 2. The domain is partitioned into 4 blocks in Level 2 and the dash lines show one of blocks in the decomposition.
(c): The DOFs in the Level 5/2. The domain is partitioned into 4 Voronoi cells about the edge centers. 
(d): The DOFs in the Level 3. This is the top level.
}
\label{fig:lel2-0}
\end{figure}

%
%
\subsubsection{Level $\ell=3/2$}
\label{sec:schur3/2}
At this level,  our goal is to find $G_{1}$ in \eqref{eqn:G1}, which is defined on the index set $J_{1}$  corresponding to boundary points of domain blocks in the first level (i.e., the black points in Figure \ref{fig:lel1-0}). We will apply Lemma \ref{lemma:4} to skeletonize the fronts. Partition the domain $\boldsymbol{\Omega}$ into $2^{L-\ell+\frac{3}{2}}(2^{L-\ell+\frac{1}{2}}-1)$ Voronoi cells \cite{Aurenhammer:1991} about the edge centers (red points in Figure \ref{fig:lel2-0} (a)). In this example, there are $24$ Voronoi cells in total since $L=3$. In Figure \ref{fig:lel2-0} (a), a Voronoi cell is an area centered at a redpoint with dashed lined compassed. Each DOF on the boundary between two adjacent Voronoi cells is randomly assigned to one and only one of these two cells. Thus a Voronoi cell includes the DOFs inside the corresponding area and some of the DOFs on its boundary. Since the DOFs of a Voronoi cell only interact with the DOFs of a few other cells nearby, that is, the matrix allows low-rank off-diagonal blocks. We can apply an ID to select the redundant and skeleton DOFs approximately in each cell and record the interpolation matrix $T_{q}$ as in Lemma \ref{lemma:2}. In the $i$th cell (see Figure \ref{fig:lel2-0} (a)), the redundant DOFs (gray points) are denoted by $I_{\frac{3}{2};i}$, the skeleton DOFs (black points) are denoted by $J_{\frac{3}{2};i}$, and the corresponding interpolation matrix is denoted by $T_{\frac{3}{2};i}$.
As in the former level, we reindex $J_{1}$ by a permutation matrix $P_{\frac{3}{2}}$ such that

\begin{equation*}\numberwithin{equation}{section}
J_{1} \stackrel{P_{\frac{3}{2}}}{\longrightarrow} (I_{\frac{3}{2};1}I_{\frac{3}{2};2}...I_{\frac{3}{2};24}|J_{\frac{3}{2};1}J_{\frac{3}{2};2}...J_{\frac{3}{2};24})=(I_{\frac{3}{2}}|J_{\frac{3}{2}}).
\end{equation*}

Denote

\begin{equation*}\numberwithin{equation}{section}
A_{\frac{3}{2}} = P_{\frac{3}{2}}^{-1}(W_{1}-V_{1}U_{1}^{-1}V_{1}^{T})P_{\frac{3}{2}} =
\left[\begin{array}{cc}
U_{\frac{3}{2}} & V_{\frac{3}{2}}^{T} \\
V_{\frac{3}{2}} & W_{\frac{3}{2}}
\end{array}\right]
\end{equation*}
with

\begin{equation*}\numberwithin{equation}{section}
U_{\frac{3}{2}} = A_{\frac{3}{2}}(I_{\frac{3}{2}},I_{\frac{3}{2}}), \quad V_{\frac{3}{2}} = A_{\frac{3}{2}}(J_{\frac{3}{2}},I_{\frac{3}{2}}), \quad \text{and}\quad W_{\frac{3}{2}} = A_{\frac{3}{2}}(J_{\frac{3}{2}},J_{\frac{3}{2}}).
\end{equation*}

Arrange $T_{1;1}$, $\dots$, $T_{1;24}$ in a block diagonal matrix

\begin{equation*}\numberwithin{equation}{section}
\label{eq:T3/2}
T_{\frac{3}{2}} =
\left[\begin{array}{ccc}
-T_{\frac{3}{2};1} & & \\
& \ddots & \\
& & -T_{\frac{3}{2};24}
\end{array}\right],
\end{equation*}
and construct a $|J_{1}|\times |J_{1}|$ matrix

\begin{equation*}\numberwithin{equation}{section}
\label{eq:Q3/2}
Q_{\frac{3}{2}} =
\left[\begin{array}{cc}
I &  \\
T_{\frac{3}{2}}& I  \\
\end{array}\right].
\end{equation*}
Then we have

\begin{equation*}\numberwithin{equation}{section}
\bar{A}_{\frac{3}{2}} = Q_{\frac{3}{2}}^{T}A_{\frac{3}{2}}Q_{\frac{3}{2}} =
\left[\begin{array}{cc}
\bar{U}_{\frac{3}{2}} & \bar{V}_{\frac{3}{2}}^{T} \\
\bar{V}_{\frac{3}{2}} & W_{\frac{3}{2}}
\end{array}\right],
\end{equation*}
where $\bar{U}_{\frac{3}{2}}$ and $\bar{V}_{\frac{3}{2}}$ are block diagonal matrices with

\begin{equation*}\numberwithin{equation}{section}
\bar{U}_{\frac{3}{2}}(I_{\frac{3}{2};i},I_{\frac{3}{2};j}) = 0,\quad  \bar{V}_{\frac{3}{2}}(J_{\frac{3}{2};i},I_{\frac{3}{2};j}) = 0, \quad\forall i \neq j.
\end{equation*}
Then

\begin{equation*}\numberwithin{equation}{section}
\bar{A}_{\frac{3}{2}}^{-1} =
\left[\begin{array}{cc}
\bar{U}_{\frac{3}{2}} & \bar{V}_{\frac{3}{2}}^{T} \\
\bar{V}_{\frac{3}{2}} & W_{\frac{3}{2}}
\end{array}\right]^{-1} =
L_{\frac{3}{2}}^{T}
\left[\begin{array}{cc}
\bar{U}_{\frac{3}{2}}^{-1} &  \\
 & G_{\frac{3}{2}}
\end{array}\right]L_{\frac{3}{2}}
\end{equation*}
with

\begin{equation*}\numberwithin{equation}{section}
L_{\frac{3}{2}}=
\left[\begin{array}{cc}
I &  \\
-\bar{V}_{\frac{3}{2}}\bar{U}_{\frac{3}{2}}^{-1} & I
\end{array}\right],\quad
G_{\frac{3}{2}} = (W_{\frac{3}{2}}-\bar{V}_{\frac{3}{2}}\bar{U}_{\frac{3}{2}}^{-1}\bar{V}_{\frac{3}{2}}^{T})^{-1}
\end{equation*}
as in Lemma \ref{lemma:4}. Note that $-\bar{V}_{\frac{3}{2}}\bar{U}_{\frac{3}{2}}^{-1}$ and $\bar{V}_{\frac{3}{2}}\bar{U}_{\frac{3}{2}}^{-1}\bar{V}_{\frac{3}{2}}^{T}$ are block diagonal.
Therefore,

\begin{equation*}\numberwithin{equation}{section}
G_{1} \approx P_{\frac{3}{2}}Q_{\frac{3}{2}}L_{\frac{3}{2}}^{T}
\left[\begin{array}{cc}
\bar{U}_{\frac{3}{2}}^{-1} &  \\
 & G_{\frac{3}{2}}
\end{array}\right]L_{\frac{3}{2}}Q_{\frac{3}{2}}^{T}P_{\frac{3}{2}}^{-1}.
\end{equation*}

Therefore, we reduce the inversion problem to a smaller matrix  $W_{\frac{3}{2}}-\bar{V}_{\frac{3}{2}}\bar{U}_{\frac{3}{2}}^{-1}\bar{V}_{\frac{3}{2}}^{T}$ by eliminating the redundant DOFs as in Lemma \ref{lemma:4}.
\subsubsection{Level $\ell=2$}
At Level $\ell=2$, the domain $\boldsymbol{\Omega}$ is partitioned into $2^{L-\ell}\times 2^{L-\ell}=2\times 2$ blocks with interior and boundary points as shown in Figure \ref{fig:lel2-0} (b). Similarly, we reindex the points in $J_{\frac{3}{2}}$ into $I_{2}$ and $J_{2}$, by a permutation matrix $P_{2}$ such that
\begin{equation*}\numberwithin{equation}{section}
J_{\frac{3}{2}} \stackrel{P_{2}}{\longrightarrow} (I_{2;11}I_{2;12}I_{2;21}I_{2;22}|J_{2;11}J_{2;12}J_{2;21}J_{2;22}):=(I_{2}|J_{2}).
\end{equation*}
Apply a similar procedure as at Level $1$ and denote

\begin{equation*}\numberwithin{equation}{section}
A_{2} = P_{2}^{-1}(W_{\frac{3}{2}}-\bar{V}_{\frac{3}{2}}\bar{U}_{\frac{3}{2}}^{-1}\bar{V}_{\frac{3}{2}}^{T})P_{2} =
\left[\begin{array}{cc}
U_{2} & V_{2}^{T} \\
V_{2} & W_{2}
\end{array}\right]
\end{equation*}
with

\begin{equation*}\numberwithin{equation}{section}
U_{2} = A_{2}(I_{2},I_{2}),\quad V_{2} = A_{2}(J_{2},I_{2}), \quad\text{and}\quad W_{2} = A_{2}(J_{2},J_{2}).
\end{equation*}
Note that $U_{2}$ and $V_{2}$ are block diagonal. Analogously,

\begin{equation*}\numberwithin{equation}{section}
G_{\frac{3}{2}} = P_{2}L_{2}^{T}
\left[\begin{array}{cc}
U_{2}^{-1} &  \\
 & G_{2}
\end{array}\right]L_{2}P_{2}^{-1},
\end{equation*}
where

\begin{equation*}\numberwithin{equation}{section}
L_{2} =
\left[\begin{array}{cc}
I &  \\
-V_{2}U_{2}^{-1} & I
\end{array}\right],\quad
G_{2} = (W_{2}-V_{2}U_{2}^{-1}V_{2}^{T})^{-1}.
\end{equation*}
Note that the update $V_{2}U_{2}^{-1}V_{2}^{T}$ is block diagonal. Now we have eliminated the interior points and the inversion problem is reduced to a smaller
matrix $W_{2}-V_{2}U_{2}^{-1}V_{2}^{T}$ as in Lemma \ref{lemma:1}.
\subsubsection{Level $\ell=5/2$}

Just as in Section \ref{sec:schur3/2}, at this level, we want to find $G_{2}$ indexed by $J_{2}$. Again, we divide the domain $\boldsymbol{\Omega}$ into $2^{L-\ell+\frac{3}{2}}(2^{L-\ell+\frac{1}{2}}-1)=4$ Voronoi cells (see Figure \ref{fig:lel2-0} (c)). The DOFs on the boundary between two cells are randomly assigned to one of cells. Through ID, we distinguish the redundant DOFs $I_{\frac{5}{2};i}$ and the skeleton DOFs $J_{\frac{5}{2};i}$ in the $i$th cell, and record the interpolation matrix $T_{\frac{5}{2};i}$. Reindexing $J_{2}$ with a permutation matrix $P_{\frac{5}{2}}$ such that

\begin{equation*}\numberwithin{equation}{section}
J_{2} \stackrel{P_{\frac{5}{2}}}{\longrightarrow} (I_{\frac{5}{2};1}I_{\frac{5}{2};2}I_{\frac{5}{2};3}I_{\frac{5}{2};4}|J_{\frac{5}{2};1}J_{\frac{5}{2};2}J_{\frac{5}{2};3}J_{\frac{5}{2};4}):=(I_{\frac{5}{2}}|J_{\frac{5}{2}}).
\end{equation*}
Denote

\begin{equation*}\numberwithin{equation}{section}
\label{eq:T5/2}
T_{\frac{5}{2}} =
\left[\begin{array}{ccc}
-T_{\frac{5}{2};1} & & \\
& \ddots & \\
& & -T_{\frac{5}{2};4}
\end{array}\right]
\end{equation*}
and a $|J_{2}|\times |J_{2}|$ matrix

\begin{equation*}\numberwithin{equation}{section}
\label{eq:Q5/2}
Q_{\frac{5}{2}} =
\left[\begin{array}{cc}
I &  \\
T_{\frac{5}{2}}& I \\
\end{array}\right].
\end{equation*}
Then

\begin{equation*}\numberwithin{equation}{section}
\bar{A}_{\frac{5}{2}} = Q_{\frac{5}{2}}^{T}P_{\frac{5}{2}}^{-1}(W_{2}-V_{2}U_{2}^{-1}V_{2}^{T})P_{\frac{5}{2}}Q_{\frac{5}{2}} =
\left[\begin{array}{cc}
\bar{U}_{\frac{5}{2}} & \bar{V}_{\frac{5}{2}}^{T} \\
\bar{V}_{\frac{5}{2}} & W_{\frac{5}{2}}
\end{array}\right]
\end{equation*}
with

\begin{equation*}\numberwithin{equation}{section}
\bar{U}_{\frac{5}{2}}(I_{\frac{5}{2};i},I_{\frac{5}{2};j}) = 0, \quad \bar{V}_{\frac{5}{2}}(J_{\frac{5}{2};i},I_{\frac{5}{2};j}) = 0,\quad  \forall i \neq j.
\end{equation*}
Therefore,

\begin{equation*}\numberwithin{equation}{section}
G_{2} \approx P_{\frac{5}{2}}Q_{\frac{5}{2}}L_{\frac{5}{2}}^{T}
\left[\begin{array}{cc}
\bar{U}_{\frac{5}{2}}^{-1} &  \\
 & G_{\frac{5}{2}}
\end{array}\right]L_{\frac{5}{2}}Q_{\frac{5}{2}}^{T}P_{\frac{5}{2}}^{-1},
\end{equation*}
where

\begin{equation*}\numberwithin{equation}{section}
L_{\frac{5}{2}} =
\left[\begin{array}{cc}
I &  \\
-\bar{V}_{\frac{5}{2}}\bar{U}_{\frac{5}{2}}^{-1} & I
\end{array}\right],\quad
G_{\frac{5}{2}} = (\bar{W}_{\frac{5}{2}}-\bar{V}_{\frac{5}{2}}\bar{U}_{\frac{5}{2}}^{-1}\bar{V}_{\frac{5}{2}}^{T})^{-1}.
\end{equation*}
Note that $U_{\frac{5}{2}}$ and $V_{\frac{5}{2}}$ are block diagonal.
The matrix inversion problem now has been reduced to $\bar{W}_{\frac{5}{2}}-\bar{V}_{\frac{5}{2}}\bar{U}_{\frac{5}{2}}^{-1}\bar{V}_{\frac{5}{2}}^{T}$.
\subsubsection{Level $\ell=3$}

The domain  $\boldsymbol{\Omega}$ is partitioned into $2^{L-\ell}\times 2^{L-\ell}=1\times 1 $ block, i.e., no partition at this level. The interior and boundary points are shown in Figure \ref{fig:lel2-0} (d). Similar to previous integer levels, reindexing $J_{\frac{5}{2}}$ into the union of an interiors index set $I_3$ and a boundary index set $J_3$ with a permutation matrix $P_3$ such that

\begin{equation*}\numberwithin{equation}{section}
J_{\frac{5}{2}} \stackrel{P_{3}}{\longrightarrow} (I_{3}|J_{3}).
\end{equation*}
Finally,

\begin{equation*}\numberwithin{equation}{section}
G_{\frac{5}{2}} = P_{3}L_{3}^{T}
\left[\begin{array}{cc}
U_{3}^{-1} &  \\
 & G_{3}
\end{array}\right]L_{3}P_{3}^{-1},
\end{equation*}
where

\begin{equation*}\numberwithin{equation}{section}
L_{3} =
\left[\begin{array}{cc}
I &  \\
-V_{3}U_{3}^{-1} & I
\end{array}\right],
G_{3} = (W_{3}-V_{3}U_{3}^{-1}V_{3}^{T})^{-1}.
\end{equation*}
We calculate the inverse of $G_{3}$ directly at this point.

\subsubsection{Summary of Construction}

We consider the construction of the hierarchical structure of Schur complements of matrix $A$ on an $N \times N$ grid. 
At each integer level, the points in each block are divided into interior points and boundary points. The interior points only interact with the points within the same block. We reindex the points and eliminate the interior points accordingly. At each fractional level, the domain is divided into Voronoi cells, and ID is applied to each unit to distinguish redundant points and skeleton points such that the redundant points only interact with the points within the same cell. We will reindex these points accordingly and eliminate the redundant points.

The following relationship is defined for each level

\begin{equation}\numberwithin{equation}{section}
\label{eq:rel}
G_{\ell} =
\begin{cases}
G = A^{-1}, & \ell = 0; \\
G_{\ell} = (W_{\ell}-V_{\ell}U_{\ell}^{-1}V_{\ell}^{T})^{-1}, & \ell \text{\ is\ integer}; \\
G_{\ell} = (W_{\ell}-\bar{V}_{\ell}\bar{U}_{\ell}^{-1}\bar{V}_{\ell}^{T})^{-1}, & \ell \text{\ is\ fractional}.
\end{cases}
\end{equation}
Based on (\ref{eq:rel}), it follows the recursive relation with integer $\ell$,

\begin{equation*}\numberwithin{equation}{section}
G_{\ell-1} \approx P_{\ell-\frac{1}{2}}Q_{\ell-\frac{1}{2}}L_{\ell-\frac{1}{2}}^{T}
\left[\begin{array}{cc}
\bar{U}_{\ell-\frac{1}{2}}^{-1} &  \\
 & G_{\ell-\frac{1}{2}}
\end{array}\right]L_{\ell-\frac{1}{2}}Q_{\ell-\frac{1}{2}}^{T}P_{\ell-\frac{1}{2}}^{-1},
\end{equation*}
\begin{equation*}\numberwithin{equation}{section}
G_{\ell-\frac{1}{2}} = P_{\ell}L_{\ell}^{T}
\left[\begin{array}{cc}
U_{\ell}^{-1} &  \\
 & G_{\ell}
\end{array}\right]L_{\ell}P_{\ell}^{-1}.
\end{equation*}
Therefore, we can construct the hierarchy of Schur complements from the bottom. We organize this algorithm in Algorithm \ref{algo1}. Note that the reindexing is implicitly included in Algorithm \ref{algo1}, when we use the index sets $I_{\ell;ij}$ and $J_{\ell;ij}$ or $I_{\ell;i}$ and $J_{\ell;i}$ for $A_{\ell}$.

\IncMargin{1em}
\begin{algorithm}
\SetAlgoNoLine
\SetKwInOut{Output}{\textbf{Output}}

\BlankLine
Determine $\ell_{\max}$ and decompose the domain hierarchically\

Generate index sets $I_{1;ij}$ and $J_{1;ij}$\

$A_{1}\leftarrow A$

\For {$\ell = 1$ to $\ell_{\max}$ }{
     $A_{\ell+\frac{1}{2}} \leftarrow A_{\ell}(J_{\ell},J_{\ell})$\

     \For {($i,j$)$\in$ $\{$block index at level $\ell$ $\}$}{
           $U_{\ell;ij} \leftarrow A_{\ell}(I_{\ell;ij},I_{\ell;ij})$\

           $V_{\ell;ij} \leftarrow A_{\ell}(J_{\ell;ij},I_{\ell;ij})$\

           Calculate $U_{\ell;ij}^{-1}$\

           Calculate $K_{\ell;ij}\leftarrow -V_{\ell;ij}U_{\ell;ij}^{-1}$\

           Calculate $A_{\ell+\frac{1}{2}}(J_{\ell;ij},J_{\ell;ij})\leftarrow A_{\ell+\frac{1}{2}}(J_{\ell;ij},J_{\ell;ij}) + K_{\ell;ij}V_{\ell;ij}^{T}$\
           }
          \If {$\ell<\ell_{\max}$}{
             Construct Voronoi cells at level  $\ell+\frac{1}{2}$\

             \For {$k$ $\in$ $\{$block index at level $\ell+\frac{1}{2}$$\}$}{
                   Use ID to compute $T_{\ell+\frac{1}{2};k}$, $I_{\ell+\frac{1}{2};k}$ and $J_{\ell+\frac{1}{2};k}$

                   $\bar{U}_{\ell+\frac{1}{2};k} \leftarrow A_{\ell+\frac{1}{2}}(I_{\ell+\frac{1}{2};k},I_{\ell+\frac{1}{2};k})$\

                   $\bar{V}_{\ell+\frac{1}{2};k} \leftarrow A_{\ell+\frac{1}{2}}(J_{\ell+\frac{1}{2};k},I_{\ell+\frac{1}{2};k})$\

                   Calculate $\bar{\ell}_{\ell+\frac{1}{2};k}\leftarrow \bar{V}_{\ell+\frac{1}{2};k}^{T}T_{\ell+\frac{1}{2};k}$\

                   Calculate $\bar{V}_{\ell+\frac{1}{2};k} \leftarrow \bar{V}_{\ell+\frac{1}{2};k}-A_{\ell+\frac{1}{2}}(J_{\ell+\frac{1}{2};k},J_{\ell+\frac{1}{2};k})T_{\ell+\frac{1}{2};k}$\

                   Calculate $\bar{U}_{\ell+\frac{1}{2};k} \leftarrow \bar{U}_{\ell+\frac{1}{2};k}- \bar{\ell}_{\ell+\frac{1}{2};k} - T_{\ell+\frac{1}{2};k}^{T}\bar{V}_{\ell+\frac{1}{2};k}$\
                   }
             $A_{\ell+1} \leftarrow A_{\ell+\frac{1}{2}}(J_{\ell+\frac{1}{2}},J_{\ell+\frac{1}{2}})$\

             \For {$k$ $\in$ $\{$block index at level $\ell+\frac{1}{2}$$\}$}{
                   Calculate $\bar{U}_{\ell+\frac{1}{2};k}^{-1}$\

                   Calculate $\bar{K}_{\ell+\frac{1}{2};k}\leftarrow -\bar{V}_{\ell+\frac{1}{2};k}\bar{U}_{\ell+\frac{1}{2};k}^{-1}$\

                   Calculate $A_{\ell+1}(J_{\ell+\frac{1}{2};k},J_{\ell+\frac{1}{2};k})\leftarrow A_{\ell+1}(J_{\ell+\frac{1}{2};k},J_{\ell+\frac{1}{2};k}) + \bar{K}_{\ell+\frac{1}{2};k}\bar{V}_{\ell+\frac{1}{2};k}^{T}$\
                  }
             Construct $I_{\ell+1}$ and $J_{\ell+1}$\
             }
     }
Calculate $G_{\ell_{\max}}\leftarrow A_{\ell_{\max}+\frac{1}{2}}^{-1}$

\Output {
        \\
        $I_{\ell},J_{\ell},I_{\ell+\frac{1}{2}},J_{\ell+\frac{1}{2}},U_{\ell;ij}^{-1},\bar{U}_{\ell+\frac{1}{2};k}^{-1},K_{\ell;ij},\bar{K}_{\ell+\frac{1}{2};k},G_{\ell_{\max}}$, for each $\ell,i,j,k$\\
}

\caption{Constructing the hierarchy of Schur complements of $A$}
\label{algo1}
\end{algorithm}
\DecMargin{1em}

\subsection{Extracting the Diagonal of the Inverse of Matrix}

After obtaining the hierarchical structure of Schur complements, we now apply the observation in Lemma \ref{lemma:1} to extract the diagonal of the inverse matrix $G$. The point is that it is not necessary to compute the whole Schur complement $G_{\ell}$. More precisely, our observations show that:

\begin{equation*}
G_{\ell-1}(I_{\ell;ij}J_{\ell;ij},I_{\ell;ij}J_{\ell;ij})\ \text{is\ determined\ by} \ G_{\ell-\frac{1}{2}}(J_{\ell;ij},J_{\ell;ij}),
\end{equation*}
\begin{equation*}
G_{\ell-\frac{1}{2}}(I_{\ell-\frac{1}{2};i}J_{\ell-\frac{1}{2};i},I_{\ell-\frac{1}{2};i}J_{\ell-\frac{1}{2};i})\  \text{is\ determined\ by}\ G_{\ell}(J_{\ell-\frac{1}{2};i},J_{\ell-\frac{1}{2};i}).
\end{equation*}
Therefore, we can develop a linear scaling algorithm to exact the diagonal elements of $G$ recursively. We organize this algorithm in Algorithm \ref{algo2}. Note that the reindexing is implicitly included in Algorithm \ref{algo2}, when we use the index sets $J_{\ell;ij}$ or $J_{\ell;i}$ for $G_{\ell}$.

\subsubsection{Level $\ell=3$}

We start from the top level $\ell=L=3$ to extract information of interest.  Given $G_{3}$, $G_{\frac{5}{2}}$ is obtained by the following formula:

\begin{equation*}\numberwithin{equation}{section}
G_{\frac{5}{2}} = P_{3}
\left[\begin{array}{cc}
U_{3}^{-1}+U_{3}^{-1}V_{3}^{T}G_{3}V_{3}U_{3}^{-1}& -U_{3}^{-1}V_{3}^{T}G_{3}\\
-G_{3}V_{3}U_{3}^{-1} & G_{3}
\end{array}\right]P_{3}^{-1}.
\end{equation*}

Submatrices in the bracket are indexed by $(I_{3}|J_{3})$. $G_{\frac{5}{2}}$ is indexed by $J_{\frac{5}{2}} = J_{\frac{5}{2};1}J_{\frac{5}{2};2}J_{\frac{5}{2};3}J_{\frac{5}{2};4}$  due to the permutation matrix $P_{3}$.
In fact, we only need to focus on $G_{\frac{5}{2}}(J_{\frac{5}{2};i},J_{\frac{5}{2};i})$ instead of off-diagonal blocks in order to extract the diagonal entries of $G_{\frac{5}{2}}$. Hence, represent $G_{\frac{5}{2}}$ as

\begin{equation*}\numberwithin{equation}{section}
G_{\frac{5}{2}} =
\left[\begin{array}{cccc}
G_{\frac{5}{2};1} & * & * & * \\
* & G_{\frac{5}{2};2} & * & * \\
* & * & G_{\frac{5}{2};3} & * \\
* & * & * & G_{\frac{5}{2};4}
\end{array}\right]
\end{equation*}
with

\begin{equation*}
G_{\frac{5}{2};i} = G_{\frac{5}{2}}(J_{\frac{5}{2};i},J_{\frac{5}{2};i}).
\end{equation*}

\subsubsection{Level $\ell=5/2$}

At Level $\ell=5/2$, we now have
\begin{equation}\numberwithin{equation}{section}
\label{eq:G2e}
G_{2} \approx P_{\frac{5}{2}}
\left[\begin{array}{cc}
\mathcal{G}_{2} & -\bar{U}_{\frac{5}{2}}^{-1}\bar{V}_{\frac{5}{2}}^{T}G_{\frac{5}{2}} + \mathcal{G}_{2}T_{\frac{5}{2}}^{T}\\
-G_{\frac{5}{2}}\bar{V}_{\frac{5}{2}}\bar{U}_{\frac{5}{2}}^{-1} + T_{\frac{5}{2}}\mathcal{G}_{2} & \mathfrak{G}_{2}
\end{array}\right]P_{\frac{5}{2}}^{-1}
\end{equation}
where

\begin{equation*}\numberwithin{equation}{section}
\mathcal{G}_{2}=\bar{U}_{\frac{5}{2}}^{-1}+\bar{U}_{\frac{5}{2}}^{-1}\bar{V}_{\frac{5}{2}}^{T}G_{\frac{5}{2}}\bar{V}_{\frac{5}{2}}\bar{U}_{\frac{5}{2}}^{-1},
\end{equation*}
and

\begin{equation*}\numberwithin{equation}{section}
\mathfrak{G}_{2} = T_{\frac{5}{2}}\mathcal{G}_{2}T_{\frac{5}{2}}^{T}-G_{\frac{5}{2}}\bar{V}_{\frac{5}{2}}\bar{U}_{\frac{5}{2}}^{-1}T_{\frac{5}{2}}^{T}-T_{\frac{5}{2}}\bar{U}_{\frac{5}{2}}^{-1}\bar{V}_{\frac{5}{2}}^{T}G_{\frac{5}{2}}+G_{\frac{5}{2}}.
\end{equation*}
Note that $T_{\frac{5}{2}}$, $U_{\frac{5}{2}}^{-1}$, and $V_{\frac{5}{2}}$ are block diagonal. We have

\begin{equation*}\numberwithin{equation}{section}
\begin{split}
\bar{U}_{\frac{5}{2}}^{-1}\bar{V}_{\frac{5}{2}}^{T}G_{\frac{5}{2}}\bar{V}_{\frac{5}{2}}\bar{U}_{\frac{5}{2}}^{-1} =
\left[\begin{array}{ccc}
\bar{U}_{\frac{5}{2};1}^{-1}\bar{V}_{\frac{5}{2};1}^{T}G_{\frac{5}{2};1}\bar{V}_{\frac{5}{2};1}\bar{U}_{\frac{5}{2};1}^{-1}& \cdots & *\\
\vdots & \ddots & \vdots\\
* & \cdots & \bar{U}_{\frac{5}{2};4}^{-1}\bar{V}_{\frac{5}{2};4}^{T}G_{\frac{5}{2};4}\bar{V}_{\frac{5}{2};4}\bar{U}_{\frac{5}{2};4}^{-1}
\end{array}\right],
\end{split}
\end{equation*} as well as

\begin{equation*}\numberwithin{equation}{section}
\begin{split}
G_{\frac{5}{2}}\bar{V}_{\frac{5}{2}}\bar{U}_{\frac{5}{2}}^{-1}T_{\frac{5}{2}}^{T} =
\left[\begin{array}{ccc}
G_{\frac{5}{2};1}\bar{V}_{\frac{5}{2};1}\bar{U}_{\frac{5}{2};1}^{-1}T_{\frac{5}{2};1}^{T}& \cdots & *\\
\vdots & \ddots & \vdots\\
* & \cdots & G_{\frac{5}{2};4}\bar{V}_{\frac{5}{2};4}\bar{U}_{\frac{5}{2};4}^{-1}T_{\frac{5}{2};4}^{T}
\end{array}\right].
\end{split}
\end{equation*}
Therefore, the corresponding diagonal blocks of $\mathcal{G}_{2}$ can be computed just using block-block multiplication accordingly.
Furthermore, similar operations can be applied to $\mathfrak{G}_{2}$.

All matrices in the bracket of (\ref{eq:G2e}) are indexed by $(I_{\frac{5}{2}}|J_{\frac{5}{2}})$. $G_{\frac{5}{2}}$ is indexed by $J_{2} = J_{2;11}J_{2;12}J_{2;21}J_{2;22}$  due to the permutation matrix $P_{\frac{5}{2}}$. 
Similar to the previous level, we only need to seek the diagonal blocks $G_{2}(J_{2;ij},J_{2;ij})$.

\IncMargin{1em}
\begin{algorithm}
\SetAlgoNoLine
\SetKwInOut{Input}{\textbf{Input}}

\BlankLine

\Input{
        \\
        Output of Algorithm \ref{algo1}\\}
\For {$\ell = \ell_{\max}$ to $1$}{
      \For {($i,j$)$\in$ $\{$block index at level $\ell$ $\}$}{
            Calculate $G_{\ell-\frac{1}{2}}(I_{\ell;ij},I_{\ell;ij})\leftarrow U_{\ell;ij}^{-1}+K_{\ell;ij}^{T}G_{\ell}(J_{\ell;ij},J_{\ell;ij})K_{\ell;ij}$\

            Calculate $G_{\ell-\frac{1}{2}}(J_{\ell;ij},I_{\ell;ij})\leftarrow G_{\ell}(J_{\ell;ij},J_{\ell;ij})K_{\ell;ij}$\

            $G_{\ell-\frac{1}{2}}(I_{\ell;ij},J_{\ell;ij})\leftarrow G_{\ell-\frac{1}{2}}(J_{\ell;ij},I_{\ell;ij})^{T}$\

            $G_{\ell-\frac{1}{2}}(J_{\ell;ij},J_{\ell;ij})\leftarrow G_{\ell}(J_{\ell;ij},J_{\ell;ij})$

      }

      \If {$\ell>1$}{
          \For {$k$ $\in$ $\{$block index at level $\ell-\frac{1}{2}$ $\}$}{
                Calculate $G_{\ell-1}(I_{\ell-\frac{1}{2};k},I_{\ell-\frac{1}{2};k})\leftarrow \bar{U}_{\ell-\frac{1}{2};k}^{-1}+\bar{K}_{\ell-\frac{1}{2};k}^{T}G_{\ell-\frac{1}{2}}(J_{\ell-\frac{1}{2};k},J_{\ell-\frac{1}{2};k})\bar{K}_{\ell-\frac{1}{2};k}$\

                Calculate $\bar{W}_{\ell-1}(J_{\ell-\frac{1}{2};k},I_{\ell-\frac{1}{2};k})\leftarrow G_{\ell-\frac{1}{2}}\bar{K}_{\ell-\frac{1}{2};k}$\

                $G_{\ell-1}(J_{\ell-\frac{1}{2};k},I_{\ell-\frac{1}{2};k})\leftarrow \bar{W}_{\ell-1}(J_{\ell-\frac{1}{2};k},I_{\ell-\frac{1}{2};k})+T_{\ell-\frac{1}{2};k}G_{\ell-1}(I_{\ell-\frac{1}{2};k},I_{\ell-\frac{1}{2};k})$\

                $G_{\ell-1}(I_{\ell-\frac{1}{2};k},J_{\ell-\frac{1}{2};k})\leftarrow G_{\ell-1}(J_{\ell-\frac{1}{2};k},I_{\ell-\frac{1}{2};k})^{T}$\

                $G_{\ell-1}(J_{\ell-\frac{1}{2};k},J_{\ell-\frac{1}{2};k})\leftarrow G_{\ell-1}(J_{\ell-\frac{1}{2};k},I_{\ell-\frac{1}{2};k})T_{\ell-\frac{1}{2};k}^{T}+T_{\ell-\frac{1}{2};k}\bar{W}_{\ell-1}(J_{\ell-\frac{1}{2};k},I_{\ell-\frac{1}{2};k})+G_{\ell-\frac{1}{2}}(J_{\ell-\frac{1}{2};k},J_{\ell-\frac{1}{2};k})$
          }
      }
}
\caption{Extracting the diagonal of $A^{-1}$}
\label{algo2}
\end{algorithm}
\DecMargin{1em}

\subsubsection{Level $\ell=2$}

At Level $\ell=2$, we have

\begin{equation}\numberwithin{equation}{section}\label{eq:G3/2}
G_{\frac{3}{2}} = P_{2}
\left[\begin{array}{cc}
U_{2}^{-1}+U_{2}^{-1}V_{2}^{T}G_{2}V_{2}U_{2}^{-1}& -U_{2}^{-1}V_{2}^{T}G_{2}\\
-G_{2}V_{2}U_{2}^{-1} & G_{2}
\end{array}\right]P_{2}^{-1}.
\end{equation}
Similar to Level 3, submatrices in the bracket of (\ref{eq:G3/2}) are indexed by $(I_{2}|J_{2})$. $G_{\frac{3}{2}}$ is indexed by $J_{\frac{3}{2}} = J_{\frac{3}{2};1}\cdots J_{\frac{3}{2};24}$ due to the permutation matrix $P_2$. Again, only $G_{\frac{3}{2}}(J_{\frac{3}{2};i},J_{\frac{3}{2};i})$ needs to be computed.

\subsubsection{Level $\ell=3/2$}

Proceeding to Level $3/2$, now

\begin{equation}\numberwithin{equation}{section}
\label{eq:G1e}
G_{1} \approx P_{\frac{3}{2}}
\left[\begin{array}{cc}
\mathcal{G}_{1} & -\bar{U}_{\frac{3}{2}}^{-1}\bar{V}_{\frac{3}{2}}^{T}G_{\frac{3}{2}} + \mathcal{G}_{1}T_{\frac{3}{2}}^{T}\\
-G_{\frac{3}{2}}\bar{V}_{\frac{3}{2}}\bar{U}_{\frac{3}{2}}^{-1} + T_{\frac{3}{2}}\mathcal{G}_{1} & \mathfrak{G}_{1}
\end{array}\right]P_{\frac{3}{2}}^{-1},
\end{equation}
where

\begin{equation*}\numberwithin{equation}{section}
\mathcal{G}_{1}=\bar{U}_{\frac{3}{2}}^{-1}+\bar{U}_{\frac{3}{2}}^{-1}\bar{V}_{\frac{3}{2}}^{T}G_{\frac{3}{2}}\bar{V}_{\frac{3}{2}}\bar{U}_{\frac{3}{2}}^{-1},
\end{equation*}
and

\begin{equation*}\numberwithin{equation}{section}
\mathfrak{G}_{1} = T_{\frac{3}{2}}\mathcal{G}_{1}T_{\frac{3}{2}}^{T}-G_{\frac{3}{2}}\bar{V}_{\frac{3}{2}}\bar{U}_{\frac{3}{2}}^{-1}T_{\frac{3}{2}}^{T}-T_{\frac{3}{2}}\bar{U}_{\frac{3}{2}}^{-1}\bar{V}_{\frac{3}{2}}^{T}G_{\frac{3}{2}}+G_{\frac{3}{2}}.
\end{equation*}
Similar to Level $\ell=5/2$, diagonal blocks of $\mathcal{G}_{2}$ and $\mathfrak{G}_{2}$ can be computed quickly using block-block multiplication accordingly. Submatrices in the bracket of (\ref{eq:G1e}) are indexed by  $(I_{\frac{3}{2}}|J_{\frac{3}{2}})$. $G_1$ is indexed by $J_{1} = J_{1;11}J_{1;12}\cdots J_{1;44}$ due to the permutation matrix $P_{\frac{3}{2}}$. Again, only the diagonal blocks $G_{1}(J_{1;ij},J_{1;ij})$ are needed.

\subsubsection{Level $\ell=1$}

At Level $1$, the same procedure is done as at  Level $2$ and Level $3$. We can obtain $G_{1}(J_{1;ij},J_{1;ij})$  from Level $\frac{3}{2}$ and $G(J_{0;ij},J_{0;ij})$ is computed directly.  Finally, the diagonal elements in $G$ can be obtained by combining the diagonal elements of each level.

\subsection{Complexity Estimate}

We next investigate the computation complexity of the SelInvHIF.  Let us assume the domain contains $N = \sqrt{N}\times \sqrt{N}$
points and  set $\sqrt{N}=2^{L}$ with $\ell_{\max}<L$.\\
We denote the number of blocks at level $\ell$ as $n_{B}(\ell)$, and the following formula holds

\begin{equation*}\numberwithin{equation}{section}
n_{B}(\ell) =
\begin{cases}
2^{2(\ell_{\max}-\ell)},& \text { $\ell$  is integer;} \\
2^{2 \ell_{\max}-2(\ell-1)}-2^{\ell_{\max}-\ell+\frac{3}{2}},& \text { $\ell$  is fractional.}
\end{cases}
\end{equation*}
The number of points of each block or cell is denoted as $n_{P}(\ell).$ Note that interior or redundant points of the previous level are not counted because they have been eliminated in previous levels. To approximate $n_{P}(\ell)$ , we use the assumption about the skeletonization in \cite{hifde}. Then it can be shown that the typical skeleton size of a cell is

\begin{equation*}\numberwithin{equation}{section}
k_{\ell} = O(\ell).
\end{equation*}
Then we have

\begin{equation*}\numberwithin{equation}{section}
n_{P}(\ell) =
\begin{cases}
2^{2(L-\ell_{\max}+1)},& \ell = 1;\\
O(2^{L-\ell_{\max}}), & \ell = \frac{3}{2};\\
O(\ell), & \ell > \frac{3}{2}.
\end{cases}
\end{equation*}

Firstly, the construction step is considered and the following steps are shown in Algorithm \ref{algo1}. At an integer level $\ell$, we need to compute the inverse of  $U_{\ell;ij}$ (Step $9$) for each block.  Then multiply the inverse with $V_{\ell;ij}$ to obtain $K_{\ell;ij}$ (Step $10$) and finally update the new $A_{\ell+\frac{1}{2}}(J_{\ell;ij},J_{\ell;ij})$ (Step $11$).  At a fractional level $\ell+\frac{1}{2}$, for each cell, we need to compute $T_{\ell+\frac{1}{2};k}$ using ID (Step $16$, since each cell only interact with $O(1)$ cells, then the cost for this step is $O(n_{P}(\ell)^{3})$). Then apply it (Step $19$, $20$, and $21$) and multiply the inverse of $\bar{U}_{\ell+\frac{1}{2};k}$ (Step $25$) with $\bar{V}_{\ell+\frac{1}{2};k}$ to get $\bar{K}_{\ell+\frac{1}{2};k}$ (Step $26$). Finally, update $A_{\ell+1}(J_{\ell+\frac{1}{2};k},J_{\ell+\frac{1}{2};k})$ (Step $27$). The computational cost for these steps at each level is $O(n_{P}(\ell)^{3})$. Furthermore, the total cost for level $\ell$ is $O(n_{B}(\ell)n_{P}(\ell)^{3})$ for $\ell>\frac{3}{2}$, since there are $n_{B}(\ell)$ blocks at Level $\ell$.

Since
 
\begin{equation*}\numberwithin{equation}{section}\label{ineq:complexity}
\begin{split}
& 2^{2(\ell_{\max}-1)} 2^{6(L-\ell_{\max}+1)}+2^{2 \ell_{\max}-1}2^{3L-3 \ell_{\max}}+\sum\limits_{\ell=2,\frac{5}{2}}^{\ell_{\max}}(n_{B}(\ell)n_{P}(\ell)^{3}) \\
& \leq C(2^{2(\ell_{\max}-1)2^{6(L-\ell_{\max}+1)}}+2^{2 \ell_{\max}-1}2^{3L-3 \ell_{\max}}+\sum\limits_{\ell=2,\frac{5}{2}}^{\ell_{\max}}(2^{2 \ell_{\max}-2\ell}\ell^{3})) \\
& \leq C_0(2^{6L-4 \ell_{\max}}+2^{3L-\ell_{\max}}+2^{2 \ell_{\max}}),
\end{split}
\end{equation*}
where $C$ and $C_0$ are constant. Let $\ell_{\max}=O(L)$, the total computational cost for the construction step is $O(N)$ (the cost for Step $32$ is $O(n_{P}(\ell)^{3})$).

Furthermore, the extraction step is analyzed now and the following steps are considered in Algorithm \ref{algo2}. At an integer level $\ell$,  $G_{\ell-\frac{1}{2}}(I_{\ell;ij},I_{\ell;ij})$ (Step $3$) and $G_{\ell-\frac{1}{2}}(J_{\ell;ij},I_{\ell;ij})$ (Step $4$) are calculated for each block. At a fractional level $\ell-\frac{1}{2}$, for each cell, we need to calculate $G_{\ell-1}(I_{\ell-\frac{1}{2};k},I_{\ell-\frac{1}{2};k})$ (Step $10$), $G_{\ell-1}(J_{\ell-\frac{1}{2};k},I_{\ell-\frac{1}{2};k})$ (Step $12$) and $G_{\ell-1}(J_{\ell-\frac{1}{2};k},J_{\ell-\frac{1}{2};k})$ (Step $14$). The computational cost for these steps at each level is $O(n_{P}(\ell)^{3})$. Hence, the total cost for level $\ell$ is $O(n_{B}(\ell)n_{P}(\ell)^{3})$ for $\ell>\frac{3}{2}$, since there are $n_{B}(\ell)$ blocks at Level $\ell$. Similarly, the complexity for the extraction step is also $O(N)$.

Therefore, the total computational complexity is $O(N)$ by combining the construction step and extraction step if the assumption in \cite{hifde} holds.

\section{Numerical Results}
\label{Numerical}
We show numerical results for the MPB equations in two dimensions to verify the performance of the proposed SelInvHIF. In particular, the scaling of the computational time by SelInvHIF is concerned. We set a uniform fugacity parameter $\Lambda=0.2$ and a coupling parameter $\Xi=1$. The error criteria are set as  $10^{-8}$ for both the PB  and the self-consistent iterations. The relative precision of the ID step is $10^{-8}$ and the initial values for the potentials in the iteration are always constant $\Phi^{(0)}=0$ in our examples. It's worth noting that the choice about the accuracy of the ID step is a balance between the number of iterations and the factorization time for constructing hierarchical structure. Dirichlet boundary conditions are used for both the PB and the DH steps. The calculation is performed on a machine with Intel Xeon 2.2GHz and 2TB memory. The statistics of calculation time are averaged over five times. We show the first example about calculating the diagonals of the inverse of the elliptic differential operator before solving  MPB equations.

{\bf Example 1 (The discrete elliptic differential operator).}
We consider the diagonal part of the inverse of the discrete elliptic differential operator as the first example. Using the five-point stencil discretization,  a 5-diagonal $N \times N$ sparse matrix $D_5$ is denoted as:

\begin{equation*}
D_5=\left(\begin{array}{cccc}
M& -I & & 0 \\
-I & \ddots & \ddots & \\
& \ddots & \ddots & -I \\
0 & & -I & M
\end{array}\right), \quad M=\left(\begin{array}{cccc}
4 & -1 & & 0 \\
-1 & \ddots & \ddots & \\
& \ddots & \ddots & -1 \\
0 & & -1 & 4
\end{array}\right)
\end{equation*}
We then calculate the diagonals of inverse of matrix $D_5$ by the SelInvHIF and the reference ``exact" method in Ref. \cite{lin1}, respectively. The diagonals of $D_5^{-1}$ are set as $d_s$ and $d_e$, respectively.
In Table \ref{Table1}, the absolute error $E_a=\sqrt{\sum(d_s-d_e)^2/N}$ between the SelInvHIF method and the reference ``exact" method obtained with corresponding matrix size are displayed. 
The relative error $E_r=\|d_s-d_e\|_2 / \|d_e\|_2$ verifies the accuracy of the SelInvHIF. The two types of error show that diagonals of inverse of matrix are stable with the increasing grid size. Finally, Table \ref{Table1} also shows the computational time of the algorithm and verifies the linear scaling of the SelInvHIF.

\begin{table}[h]
\centering
\setcaptionwidth{0.8\textwidth}
\begin{tabular}{rcccc}
\toprule
\midrule
 Matrix size $\sqrt{N}$ & SelInvHIF time & $E_a$ & $E_r$ \\
\midrule
 \multicolumn{1}{c}{$256$}  & $2.23E+1$ & $2.12E - 8$& $2.37E - 8$ \\
 \multicolumn{1}{c}{$512$}  & $1.05E+2$ & $1.13E - 7$  & $1.13E - 7$ \\
 \multicolumn{1}{c}{$1024$}  & $4.91E+2$&   $3.87E - 7$ & $3.49E - 7$ \\
\midrule
\bottomrule
\end{tabular}

\caption{The CPU time, accuracy, and matrix size. The SelInvHIF time means the execution time spent for one step SelInvHIF.}

\label{Table1}
\end{table}

Now we define the relative error to measure the accuracy of the SelInvHIF in solving the following MPB equations:
 \[
e_r=\frac{\|\Phi-\Phi_{\mathrm{ref}}\|_2}{  \|\Phi_{\mathrm{ref}}\|_2  }
 \]
where $\Phi$ is the electric potential computed at $y=L/2$ using SelInvHIF and $\Phi_{\mathrm{ref}}$ is the electric potential at $y=L/2$ computed with sufficiently large grid size. To measure the accuracy of the whole algorithm and the convergence with respect to the grid size, we also compute the absolute error $e_a=\sqrt{\sum(\Phi-\Phi_{\mathrm{ref}})^2/N}$ using a reference solution $\Phi_{\mathrm{ref}}$ computed with sufficiently large grid size. For each example, the following notations are given:
\begin{itemize}
    \item[-] $t_T:$ the computational time for one step iteration in the whole program;
    \item[-] $t_f:$ the computational time for constructing the factorization $A$ in each iteration;
    \item[-] $t_e:$ the computational  time for extracting the diagonal part in each iteration;
    \item[-] $m_f:$ the required memory for the factorization $A$ in GB;
\end{itemize}

\begin{figure}[h]
\setcaptionwidth{0.8\textwidth}
	\centering
	\includegraphics[width=1\textwidth]{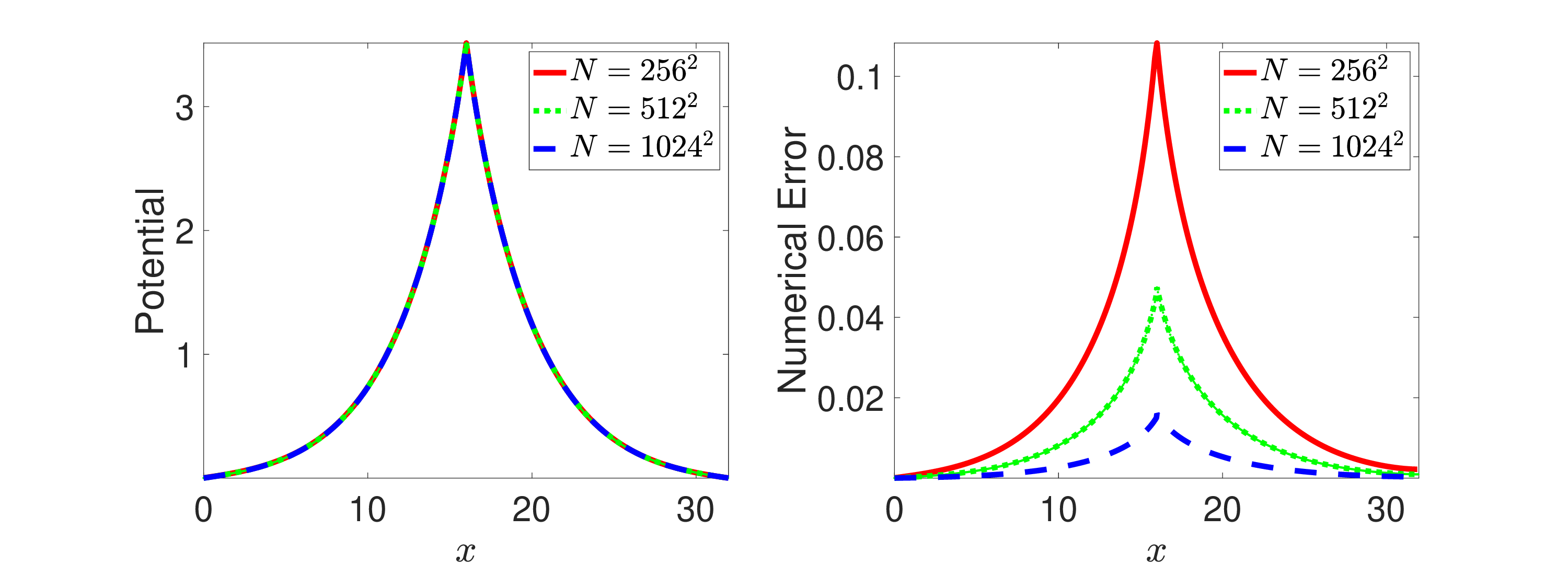}
	\caption{Numerical results about the charge density with a delta function. Left: potential distributions with different matrix size. Right:  the absolute  error between the numerical results and the reference solution with  $N =2048^2$.
}
        \label{ex:1}
\end{figure}

\begin{figure}[H]
\setcaptionwidth{0.8\textwidth}
	\centering
	\includegraphics[width=1\textwidth]{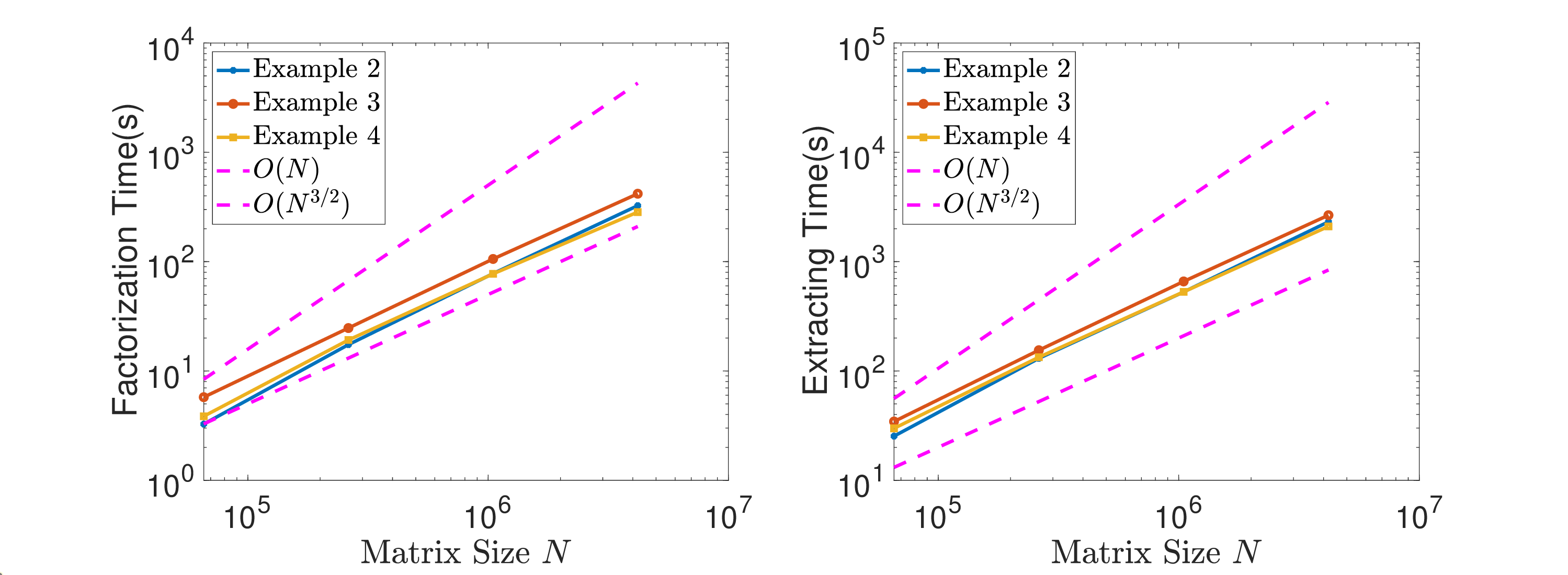}
	\caption{Scaling results for computational  time in factorization step and extracting step, respectively. The solid lines represent the computational time for one step SelInvHIF under the different  charge distribution. The reference scalings (purple dashed lines) of $O(N)$ and $O(N^{3/2})$.}
	\label{time}
\end{figure}

{\bf Example 2(The charge density with a delta function).}
We consider discontinuous charged distribution in a region $[0, L]^2$ with $L = 32$. Let the charge density be

\begin{equation*}
\rho_{f}(x)=2\delta(x-L / 2).
\end{equation*}
We then calculate the results of the MPB equations by the SelInvHIF. The left panel of Figure \ref{ex:1}  visualizes the distribution of the convergent potential at $y=L/2$ in this system with different matrix sizes $N=256^{2}$, $512^{2}$ and $1024^{2}$. The right panel of Figure \ref{ex:1} displays the numerical error $|\Phi-\Phi_{\mathrm{ref}}|$ representing 
the difference vector between the numerical results and the reference solution obtained with a sufficiently large grid size $N=2048^2$. The relative $L^2$ errors maintain approximate accuracy of first-order in Table \ref{Table2} due to the discontinuous of the derivative of the potential at $y=L/2$. Table \ref{Table2}  also shows the accuracy of the whole algorithm to compute the potential $\Phi$ compared to a reference potential computed with a sufficiently large grid size $N=2048^2$, which verifies the convergence of our algorithm. Furthermore, Table \ref{Table2} also shows the computational time of the algorithm to verify the linear scaling of SelInvHIF. Finally, the scaling results of the SelInvHIF algorithm are shown in Figure \ref{time}.

\begin{table}[h]
\centering
\setcaptionwidth{0.8\textwidth}
\begin{tabular}{rcccccc}
\toprule
\midrule
 Matrix size $\sqrt{N}$ & $t_T$ & $t_f$ & $t_e$ &$m_f$& $e_a$ & $e_r$ \\
\midrule
 \multicolumn{1}{c}{$256$}  & $3.40E+1$& $3.27E + 0$ &2.53E + 1 & $5.0E -1$  &$3.47E - 2$& $2.87E - 2$ \\
 \multicolumn{1}{c}{$512$}   & $1.82E+2$ & $1.74E + 1$ & $1.30E+2$  &  $2.0E +0$ &$1.49E - 2$ & $1.23E - 2$ \\
  \multicolumn{1}{c}{$1024$}   & $7.70E+2$ & 7.78E + 1 & $5.28E + 2$  &   $ 8.0E +0$  & $5.21E - 3$ & $4.12E - 3$  \\
   \multicolumn{1}{c}{$2048$}   & $2.77E+3$ &3.25E + 2 &$2.32E+3$ &$3.2 E +1$ & - & -  \\
\midrule
\bottomrule
\end{tabular}
\caption{The CPU time, accuracy, memory, and matrix size. }
\label{Table2}
\end{table}

\begin{figure}[h]
\setcaptionwidth{0.8\textwidth}
	\centering
	\includegraphics[width=1\textwidth]{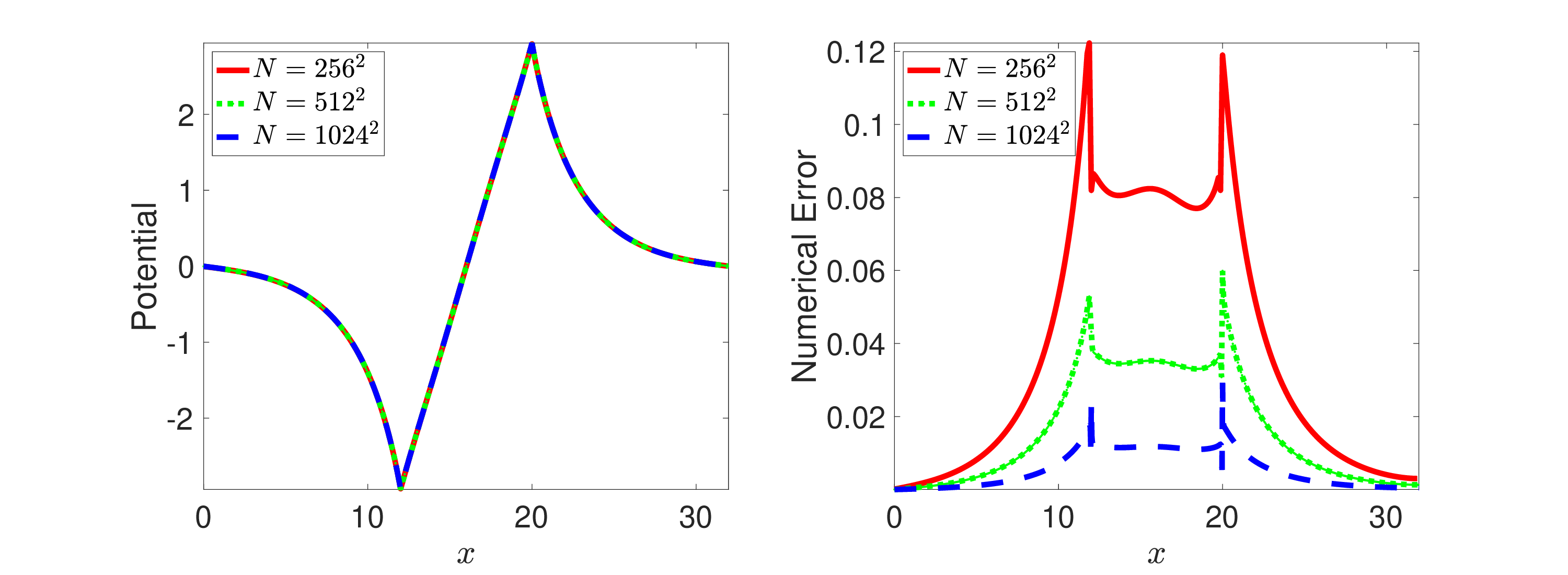}
	\caption{Numerical results about the charge density with a combined delta function. Left: potential distributions with different matrix size; Right:  the absolute  error between the numerical results and the reference solution with $N = 2048^2$ .}
	\label{ex:2}
\end{figure}

{\bf Example 3 (The charge density with radial symmetry).} In the third example, the charge density with radial symmetry is considered.  The computational interval is $[0, L]^2$ with $L = 32$ and the fixed charge density is

\begin{equation*}
\rho_{f}(x, y)=2\operatorname{sign}(x) \delta\left(\sqrt{x^{2}+y^{2}}-4\right)
\end{equation*}\\
We solve the MPB equations using the SelInvHIF. Similarly, the left panel of Figure \ref{ex:2} visualizes the distribution of the convergent potential at $y = 0.5L$ in this system with different matrix sizes $N=256^{2}$, $512^{2}$, and $1024^{2}$. The right panel of Figure \ref{ex:2} displays the numerical error $|\Phi-\Phi_{\mathrm{ref}}|$ between the numerical results and the reference solution obtained with a sufficiently large grid size $N=2048^2$. The relative $L^2$ errors maintain the approximate accuracy of first order in Table \ref{Table3} due to the discontinuity of the potential derivative at circle. Table \ref{Table3}  also shows the accuracy of the whole algorithm to compute the potential $\Phi$ compared to a reference potential, which verifies the convergence of our algorithm. Furthermore, it is shown that the computational time of the algorithm verifies the linear scaling of the SelInvHIF.  Finally, the scaling results of the SelInvHIF algorithm are shown in Figure \ref{time}.

\begin{table}[h]
\centering
\setcaptionwidth{0.8\textwidth}
\begin{tabular}{rcccccc}
\toprule
\midrule
 Matrix size $\sqrt{N}$ & $t_T$ & $t_f$ & $t_e$ &$m_f$& $e_a$ & $e_r$ \\
\midrule
 \multicolumn{1}{c}{$256$}  & $4.42E+1$ &  $5.76E + 0$ &$3.45E + 1$ & $5.0E -1$  &$5.27E - 2$ & $4.39E - 2$ \\
 \multicolumn{1}{c}{$512$}   & $1.97E+2$ & $2.47E + 1$ & $1.55E+2$  &  $2.0E +0$ & $2.27E - 2$   & $1.89E - 2$\\
  \multicolumn{1}{c}{$1024$}   & $8.37E+2$ & $1.06E + 2$ & $6.59E + 2$  &   $ 8.0E +0$  &  $7.62E-3$ & $6.32E - 3$ \\
   \multicolumn{1}{c}{$2048$}   & $3.45E+3$ &$4.18E + 2$ &$2.66E+3$ &$3.2 E +1$ & - & -  \\
\midrule
\bottomrule
\end{tabular}
\caption{The CPU time, accuracy, memory, and matrix size. }
\label{Table3}
\end{table}

\begin{figure}[h]
\setcaptionwidth{0.8\textwidth}
	\centering
	\includegraphics[width=1\textwidth]{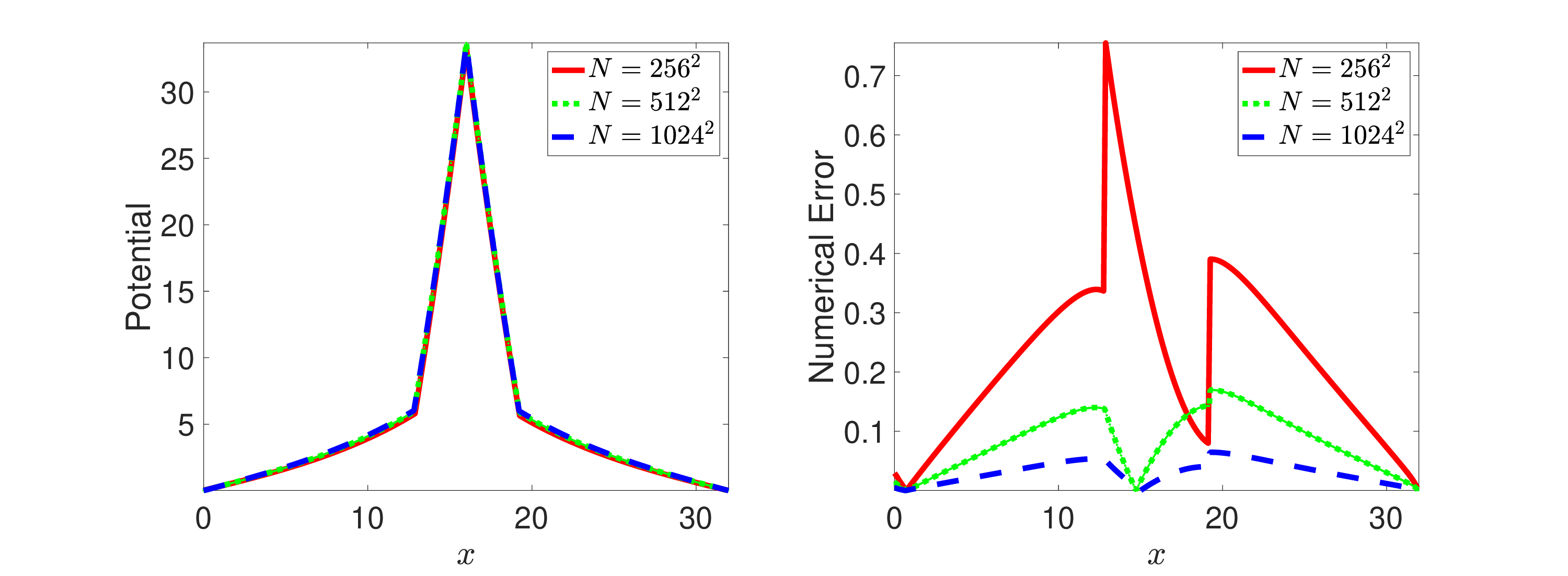}
	\caption{Numerical results about the system with dielectric discontinuity. Left: potential distributions with different matrix size; Right:  the absolute  error between the numerical results and the reference solution with $N = 2048^2$ .}
	\label{ex:3}
\end{figure}

{\bf Example 4 (The system with dielectric discontinuity).}
In the last example, we consider the system with dielectric discontinuity. The computational interval is $[0, L]^2$ with $L$ = 32, where the region of $[0.4L, 0.6L]$ is inaccessible to ions. The fixed charge density is

\begin{equation*}
\rho_{f}(x)=2\delta(x-0.5 L).
\end{equation*}
The dielectric in the region of $[0.4L, 0.6L]$  is different from the other region and the dielectric ratio is set to be $\varepsilon=0.1$.
We solve the MPB equations using the SelInvHIF. The left panel of Figure \ref{ex:3} visualizes the distribution of the convergent potential at $y=0.5L$ in this system with different matrix sizes $N=256^{2}$, $512^{2}$, and $1024^{2}$. The right panel of Figure \ref{ex:3} displays the numerical error $|\Phi-\Phi_{\mathrm{ref}}|$ between the numerical results and the reference solution obtained with a sufficiently large grid size $N=2048^2$. The relative $L^2$ errors maintain approximate accuracy of first-order in Table \ref{Table4}. Table \ref{Table4}  also shows the accuracy of the whole algorithm to compute the potential $\Phi$ compared to a reference potential, which verifies the convergence of our algorithm. Furthermore, Table \ref{Table4} also shows the computational time of the algorithm verify the linear scaling of the SelInvHIF.  Finally, the scaling results of the SelInvHIF algorithm are shown in Figure \ref{time}.

\begin{table}[h]
\centering
\setcaptionwidth{0.8\textwidth}
\begin{tabular}{rcccccc}
\toprule
\midrule
 Matrix size $\sqrt{N}$ & $t_T$ & $t_f$ & $t_e$ &$m_f$& $e_a$ & $e_r$ \\
\midrule
 \multicolumn{1}{c}{$256$}  & $4.03E+1$ &  $3.85E + 0$ &$2.90E + 1$ & $5.0E -1$  &$2.63E - 1$ & $2.78E - 2$ \\
 \multicolumn{1}{c}{$512$}   & $1.94E+2$ & $1.92E + 1$ & $1.36E+2$  &  $2.0E +0$ & $9.62E - 2$   & $9.94E - 3$\\
  \multicolumn{1}{c}{$1024$}   & $7.71E+2$ & $7.73E + 1$ & $5.29E + 2$  &   $ 8.0E +0$  &  $3.56E - 2$ & $3.36E - 3$ \\
   \multicolumn{1}{c}{$2048$}   & $2.99E+3$ &$2.84E + 2$ &$2.10E+3$ &$3.2 E +1$ & - & -  \\
\midrule
\bottomrule
\end{tabular}
\caption{The CPU time, accuracy, memory, and matrix size. }
\label{Table4}
\end{table}

\section{\large Conclusions}
\label{Conclusion}
A fast algorithm, SelInvHIF, is proposed to solve the MPB equations by combining the hierarchical interpolative factorization and the original selected inverse method. An $O(N)$ computational complexity in terms of the number of operations and memory is achieved to obtain the diagonal of the inverse of a sparse matrix discretized from an elliptic differential operator. We applied this algorithm to the two-dimensional MPB problems and attractive performance is obtained in terms of both accuracy and efficiency in solving the MPB equations. In the future, we will try to develop another fast algorithm with $O(N)$ complexity for three-dimensional problems based on a similar construction.

\section*{Acknowledgment}
Y. Tu and Z. Xu acknowledge the financial support from the National Natural Science Foundation of China (grant No. 12071288),
Science and Technology Commission of Shanghai Municipality (grant Nos. 20JC1414100 and 21JC1403700) and Strategic Priority Research Program of Chinese Academy of Sciences (grant No. XDA25010403). Q. Pang and H. Yang thank the support of the US National Science Foundation under award DMS-1945029.
\bibliographystyle{unsrt}
\bibliography{ref}

\end{document}